\documentclass{emulateapj}

\usepackage{natbib}
\usepackage{color}
\usepackage{lscape}
\newcommand{\gpos}{\centering}

\newcommand{\sly}[1]{}
\newcommand{\update}[1]{}

\newcommand{\Ha}{H$\alpha$}
\newcommand{\Msun}{M$_\odot$}

\newcommand{\etal}{et~al.}
\newcommand{\nbyso}{1657}
\newcommand{\III}{II-III}

\begin{document}
\title{The North American and Pelican Nebulae I. IRAC Observations}

\slugcomment{Version from \today}

\author{S.\ Guieu\altaffilmark{1},
  L.\ M.\ Rebull\altaffilmark{1}, 
  J.\ R.\ Stauffer\altaffilmark{1},
  L.\ A.\ Hillenbrand\altaffilmark{2},
  J.\ M.\ Carpenter\altaffilmark{2},
  A.\ Noriega-Crespo\altaffilmark{1},          
  D.\ L.\ Padgett\altaffilmark{1},
  D.\ M.\ Cole\altaffilmark{3}, 
  S.\ J.\ Carey\altaffilmark{1},
  K.\ R.\ Stapelfeldt\altaffilmark{3}, 
  S.\ E.\ Strom\altaffilmark{4}
}

\altaffiltext{1}{Spitzer Science Center/Caltech, M/S 220-6, 1200
E.\ California Blvd., Pasadena, CA  91125
(guieu@ipac.caltech.edu)}
\altaffiltext{2}{Department of Astronomy, California Institute of
Technology, Pasadena, CA 91125}
\altaffiltext{3}{Jet Propulsion Laboratory, MS 183-900, California Institute
of Technology, Pasadena, CA 91109}
\altaffiltext{4}{NOAO, Tucson, AZ}

\begin{abstract}

We present a 9 deg$^2$ map of the North American and Pelican Nebulae
regions obtained in all four IRAC channels with the Spitzer Space
Telescope. The resulting photometry is merged with that at $JHK_s$
from 2MASS and a more spatially limited $BVI$ survey from previous
ground-based work. We use a mixture of color- color diagrams to select
a minimally contaminated set of
more than 1600 objects that we claim 
are young stellar objects (YSOs) associated with the star forming
region. Because our selection technique uses IR excess as a
requirement, our sample is strongly biased against
inclusion of Class~III YSOs.
The distribution of IRAC spectral slopes  for our YSOs indicates that most 
of these objects are Class\,II, with a peak towards steeper spectral
slopes but a substantial contribution from a tail of flat spectrum and
Class\,I type objects. By studying the small fraction of the sample
that is optically visible,
we infer a typical age of a few Myr for the low mass population.
The young stars are clustered, with about a third of them
located in eight clusters that are located within or near the
LDN 935 dark cloud.  Half of the YSOs are located in regions with
surface densities higher than 1000 YSOs / deg$^2$.
The Class\,I objects are more clustered than the Class\,II stars.

\end{abstract} 

\section{Introduction}

Much of our current knowledge regarding star-forming patterns and
circumstellar disk evolution derives from study of molecular cloud
complexes within a few hundred parsecs of the Sun.  Among these 
are a large number of lower-mass clouds such as Taurus and, more
infrequently, dense clouds like Orion, which is the prototypical
high-mass and high density star forming region.  While nearby cloud
complexes serve as our primary empirical guide to understanding
the formation and early evolution of stars, 
it is
important that we study more than just the nearest examples. One
little-studied-to-date  
star-forming region beyond the solar neighborhood is the North
American and Pelican Nebulae (NGC\,7000 and IC\,5070, respectively),
towards $l$=85, $b$=$-$0.5, about 600\,pc from Earth. 
This region is probably more representative of star formation in the
disk of the Milky Way than Taurus or Orion, it is also the next
closest region after Orion 
with a substantial numbers of intermediate and high mass stars.  
The physical
appearance of the nebulae on optical images is thought to reflect a
combination of a large, background \ion{H}{2} region (W80) overlaid by
several foreground, dark clouds, with the edges of the dark clouds
illuminated by the optically-unseen primary exciting star of the
\ion{H}{2} region.  Because these nebulae lie essentially in the plane of the
Galaxy and, as seen by us, along a spiral arm, study of the region is
further confused by the juxtaposition of young stars and star-forming
regions at a variety of distances.  For the purposes of the rest of
the paper, we will refer to the entire region as the NANeb, or the NAN
complex.

\citet{herbig-1958} was one of the first to study the star formation
associated with these nebulae.  He identified 68 \Ha\ emission line
stars from objective-prism plates of the region, including a small
cluster of pre-main sequence (PMS) stars in the ``Gulf of Mexico" portion of the NANeb. 
Many of the young stars identified by Herbig lie along the bright rims
of the dark clouds, possibly indicating that they are recent, second
generation star formation products triggered by the expanding
\ion{H}{2} region from the first generation of stars in the NANeb.
Herbig estimated a rough distance of 500\,pc for the star-forming
complex, but with a variety of caveats related to whether this
distance applied to some or all of the stars and nebulae. 
The best available modern distance for the
region appears to be that of \citet{laugalys-2002}, who estimate a
distance of $\simeq$600 pc. We will use this distance for the
purposes of the paper with the caveats noted above. 

\citet{Bally-1980} mapped the CO associated with the dark clouds that
define the Gulf of Mexico and Atlantic Ocean, and estimated that the
mass in gas still present is of order $3\times10^4\,M_\odot$.  Their
interpretation of the CO data was that the remaining dark clouds are
the remnants of a much larger molecular cloud complex that has largely
been disrupted by the hot, massive stars that were formed a few
million years ago.  \citet{Cambresy-2002} used the Two-Micron All-Sky
Survey (2MASS; \citealp{Skrutskie-2006}) point-source catalog to map the
extinction and star clustering towards this region, deriving
extinctions of $A_v \geq$ 35 magnitudes in some portions of the
darkest cloud (Lynds\,935).  Cambr\'esy et al. also identified nine apparent
star clusters in the 2MASS data. 

\citet{Bally-2003} obtained narrow-band H$\alpha$ and [\ion{S}{2}]
imaging of the region of the Pelican nebula, and identified a number
of new HH objects and collimated outflows, mostly originating from the
interface region between the dark cloud and the \ion{H}{2} region.  In
a couple of cases, the flows originate from near the tip of elephant
trunks.   The exciting sources for most of the flows could not be
identified from the existing data.  \citet{armond-2003} obtained
additional narrow and broad band optical and near-IR imaging of the
NANeb, and identified 28 more HH objects, including many associated
with the PMS stars of Herbig's Gulf of Mexico cluster. 
More recently, the probable exciting source for the \ion{H}{2} region
has been identified as an O5 star behind about $A_V$=10 mag of
extinction, located inside the dark cloud LDN\,935 separating the
North America and the Pelican Nebulae \citep{Comeron-2005b}. 

We have conducted a large ($\sim$9 deg$^2$) infrared imaging survey
with Spitzer Space Telescope \citep{Werner-2004} of this region, along with supporting
data obtained in the optical for the $\sim 2.4\arcdeg \times 1.7\arcdeg$
central region.  This present paper, the first of a series, presents
the Infrared Array Camera (IRAC) data.  Future
papers will cover the Multiband Imaging Photometer for Spitzer (MIPS) 
data, with special emphasis on the Gulf of
Mexico cluster (Rebull \etal\ 2009; hereafter R09), and the optical
classification spectroscopy (Hillenbrand \etal\ in
prep). 

First, we present the observational details for the IRAC observations
(\S\ref{sec:obs}). 
We then use the IRAC colors to select a minimally contaminated, if not complete,
sample of YSO candidates (\S\ref{sec:ysos}), and discuss their
properties (classes, spatial distribution) in the context of other
star-forming regions. Because IRAC is very sensitive to emission from
HH objects, we then discuss the IRAC observations of one of the HH
objects found in this region (\S\ref{sec:hhobj}).

\section{Observations, Ancillary Data, and Basic Data Reduction}
\label{sec:obs}

\begin{deluxetable}{lccl}
\tablecaption{Summary of IRAC observations (programs 20015 and 462)\label{tab:obs}}
\tablewidth{0pt}
\tablehead{
\colhead{Program : field-ID} & \colhead{map center} &  \colhead{AORKEY} }
\startdata
P20015 : 12 & 20h55m30.00s,+44d59m00.0s & 16790528 \\
P20015 : 13 & 20h53m06.00s,+44d59m00.0s & 16790784 \\
P20015 : 21 & 20h57m54.00s,+44d23m00.0s & 16791040 \\
P20015 : 22 & 20h55m30.00s,+44d23m00.0s & 16791296 \\
P20015 : 23 & 20h53m06.00s,+44d23m00.0s & 16791552 \\
P20015 : 24 & 20h50m42.00s,+44d23m00.0s & 16791808 \\
P20015 : 31 & 20h57m54.00s,+43d47m00.0s & 16792064 \\
P20015 : 32 & 20h55m30.00s,+43d47m00.0s & 16792320 \\
P20015 : 33 & 20h53m06.00s,+43d47m00.0s & 16792576 \\
P20015 : 34 & 20h50m42.00s,+43d47m00.0s & 16792832 \\
P20015 : 42 & 20h55m30.00s,+43d11m00.0s & 16793088 \\
P20015 : 43 & 20h53m06.00s,+43d11m00.0s & 16793344 \\
P00462 : 0   & 20h50m38.00s,+45d05m10.0s   &   24251392 \\        
P00462 : 1   & 21h01m13.00s,+44d54m43.0s   &   24251648 \\        
P00462 : 2   & 20h58m51.00s,+45d12m42.0s   &   24251904 \\        
P00462 : 3   & 20h59m05.00s,+42d59m38.0s   &   24252160 \\        
P00462 : 4   & 21h00m17.00s,+43d23m25.0s   &   24252416 \\        
P00462 : 5   & 21h01m50.00s,+44d10m34.0s   &   24252672 \\        
P00462 : 6   & 20h50m19.00s,+42d58m37.0s   &   24252928 \\        
P00462 : 7   & 20h48m44.00s,+43d17m51.0s   &   24253184 \\        
\enddata
\end{deluxetable}
\subsection{Observations}

The Spitzer observations of the NAN complex were obtained as part of
the Cycle-2 program 20015 (PI: L. Rebull).  Additional observations of
the ``corners'' of the map were obtained as part of program 462 in an
effort to increase the legacy value of the data set. 
The IRAC observations from program 20015 were obtained 9-11 August
2006, amd the IRAC observations from program 462 were obtained 15-27
November 2007. IRAC \citep{Fazio-2004} observes at 3.6, 4.5, 5.8, and 8\,$\mu$m.

\begin{figure*}
\plotone{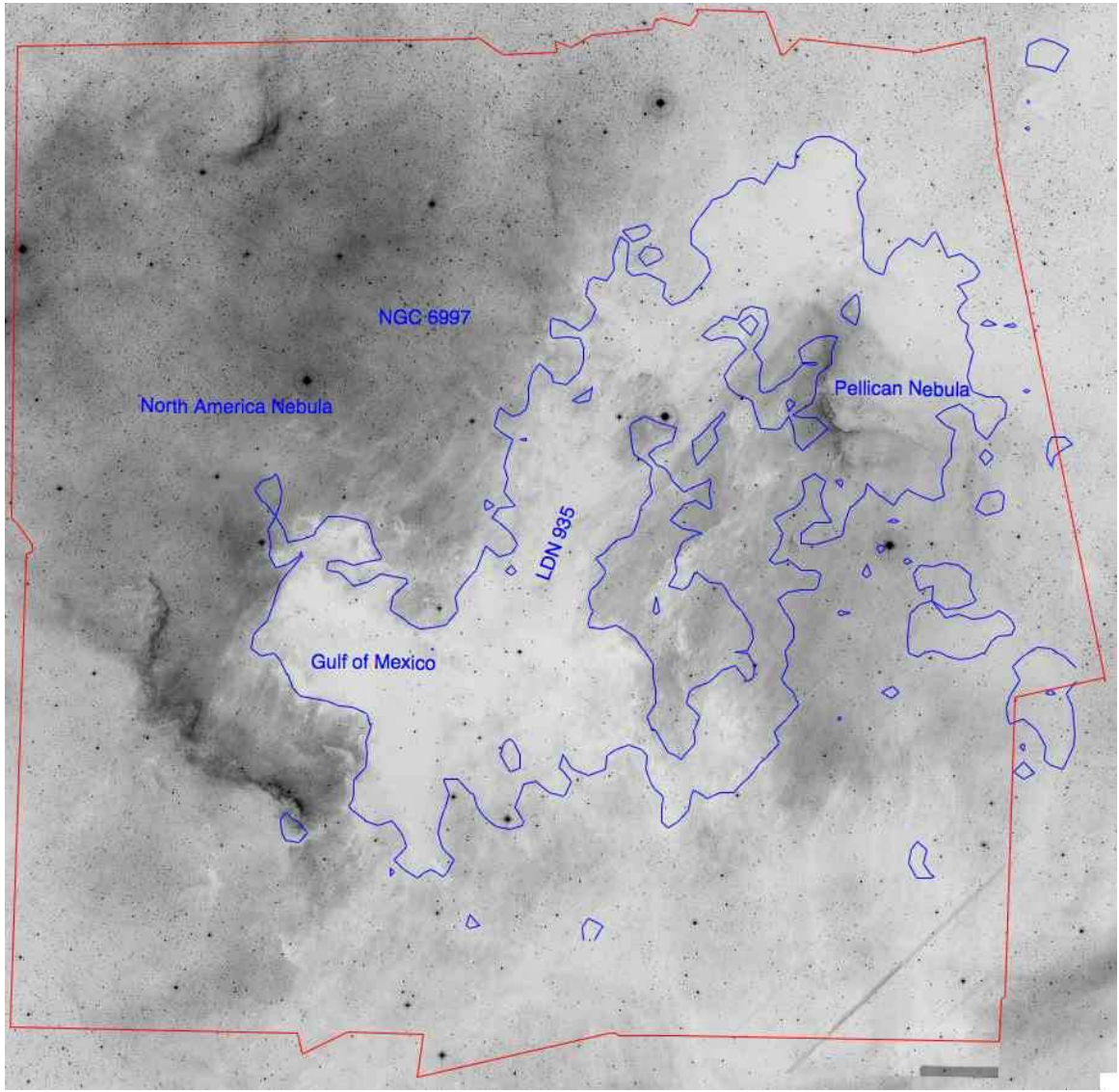}
\caption{Optical image ($\sim 4\times3.2$ degrees) of the NANeb region 
from the Palomar all-sky
survey. North is up, East is to the left. The left and right edges are
at 21h02m32s and 20h46m52s; the top and bottom edges are at 45d25m40s
and 42d40m56s. 
The region observed by IRAC is indicated by red lines. The
green line is a smoothed $A_v$=5 contour from the \cite{Cambresy-2002}
extinction map. The galactic coordinate of the center of this image is
l=84.8, b=-0.6.}
\label{fig:mosaic_optical}
\end{figure*}

The Cycle-2 IRAC observations were designed to cover the region of
highest extinction in a manner as independent of observing constraints
as possible.  Figure \ref{fig:mosaic_optical}  is the NANeb region in
the optical (from
the Digitized Palomar Optical Sky Survey).  The region covered by our IRAC
map is indicated, as is the $A_v=5$ contour from the
\cite{Cambresy-2002} extinction map. 
The observations were broken into 12
astronomical observation requests (AORs); the AORKEYs are given in
Table~\ref{tab:obs}.  This region is at a high ecliptic
latitude (+57$^{\circ}$), so the field of view rotates by about a
degree a day, necessitating small individual AOR coverage in order to
completely tile the region without leaving gaps. Also, because of
the high ecliptic latitude, asteroids are not a significant concern,
and we therefore obtained all of the imaging at a single epoch.  Each
mapping AOR was constructed with the same strategy -- at each map
step, four dither positions were observed, each with high-dynamic-range
(HDR) exposures of 0.6 and 12.0 second frame times.  
For this strategy, the on-line SENS-PET for a high-background region
reports 3-$\sigma$ point-source sensitivities of 7.2, 10.7, 66, and 78\,$\mu$Jy
for the 4 IRAC bands, respectively. 
Differential source count histograms for our point sources are
very similar to those for other star-forming regions observed in a
similar manner - see in particular Figure 8 of Winston et al. 2007
(ApJ 669, 493) for 3.6 $\mu$m; the NaNeb histograms indicate that 
our 90\% completeness limits
are about [3.6] = 15 decreasing to about [8.0] = 12.5.  For our purposes,
however, the more important number is the completeness for detecting
objects in all four channels.  We estimate this by determining where
the four-channel differential source counts drops below the 3.6 $\mu$m
source counts by more than 20\%, which occurs at about [3.6] = 12.2.
This is about as expected given that four-channel catalog is normally
limited by the detections in the 8.0 $\mu$m channel.
This completeness limit
corresponds to a spectral type of M4 or M1 
according to the BCAH98 isochrones \citep{Baraffe-1998} for 1
or 5 Myr, respectively, a distance of 600\,pc (assuming no reddening)
and the temperature to spectral type scale defined by \cite{Luhman-2003b}. 
The faintest object in our final catalog has [3.6]$\sim$16,
corresponding to a mass of 0.02-0.03\,\Msun\ at 1-5\,Myr age, assuming the 3.6\,$\mu$m 
flux is photospheric.

The remaining IRAC observations from program 462 were designed 
using the fixed-cluster observing mode 
to cover the irregularly-shaped regions to the same depth as
the main map  and to create a final map that is approximately
square in shape.  The map center given in Table~\ref{tab:obs} is
the approximate center of each AOR. We
used the software developed by R.~Gutermuth and T.~Megeath (private
communication) for use by the IRAC GTO and Gould's Belt teams to
construct these AORs. 

\subsection{Basic Data Reduction}

\begin{figure*}
\plotone{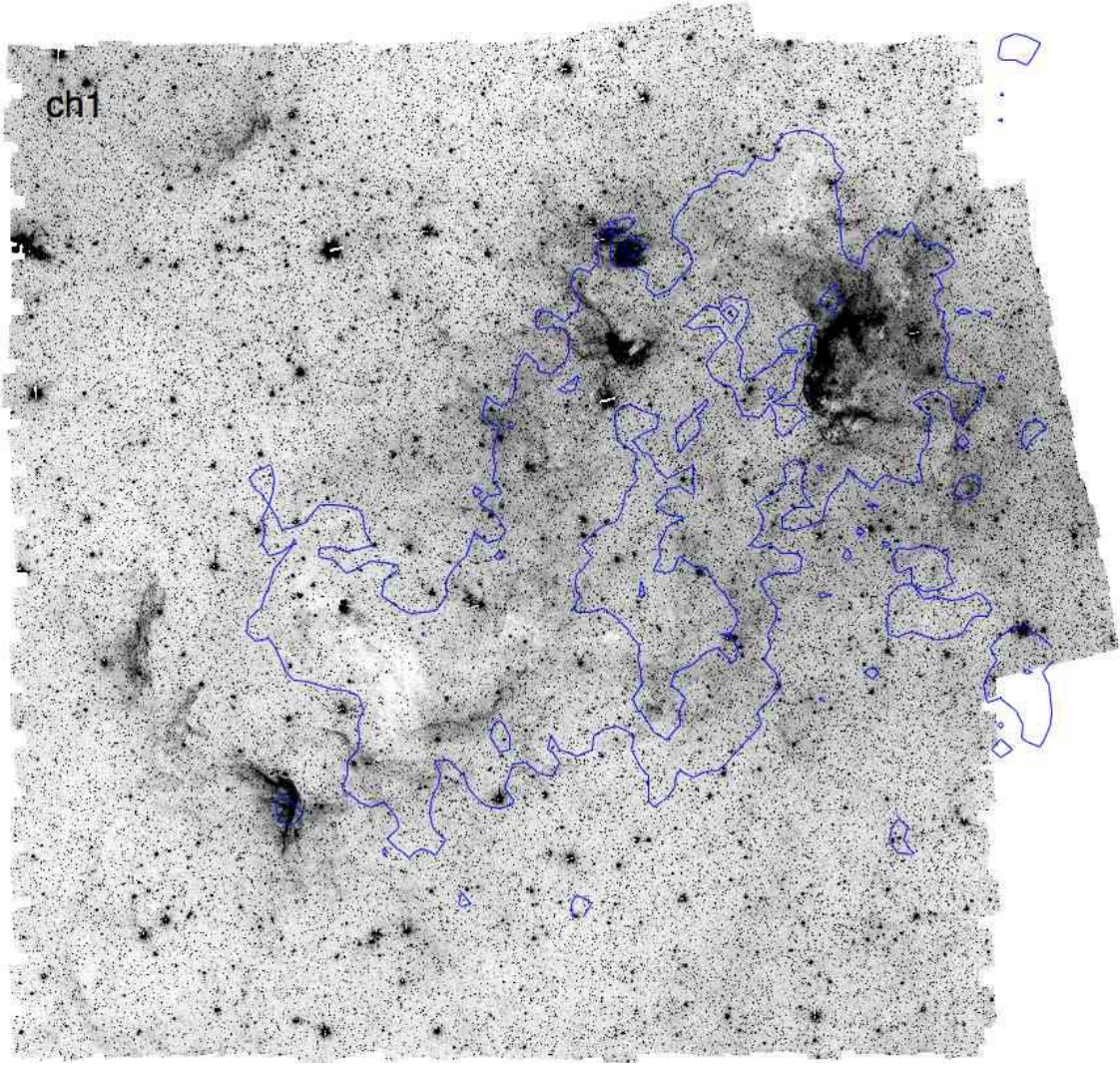}
\caption{ 
Mosaic of the NANeb region from IRAC channel 1
(3.6\,$\mu$m). Orientation and map center are the same as previous
Figure; overlays from previous Figure are included for reference.  
}
\label{fig:mosaic_I1}
\end{figure*}

\begin{figure*}\gpos
\plotone{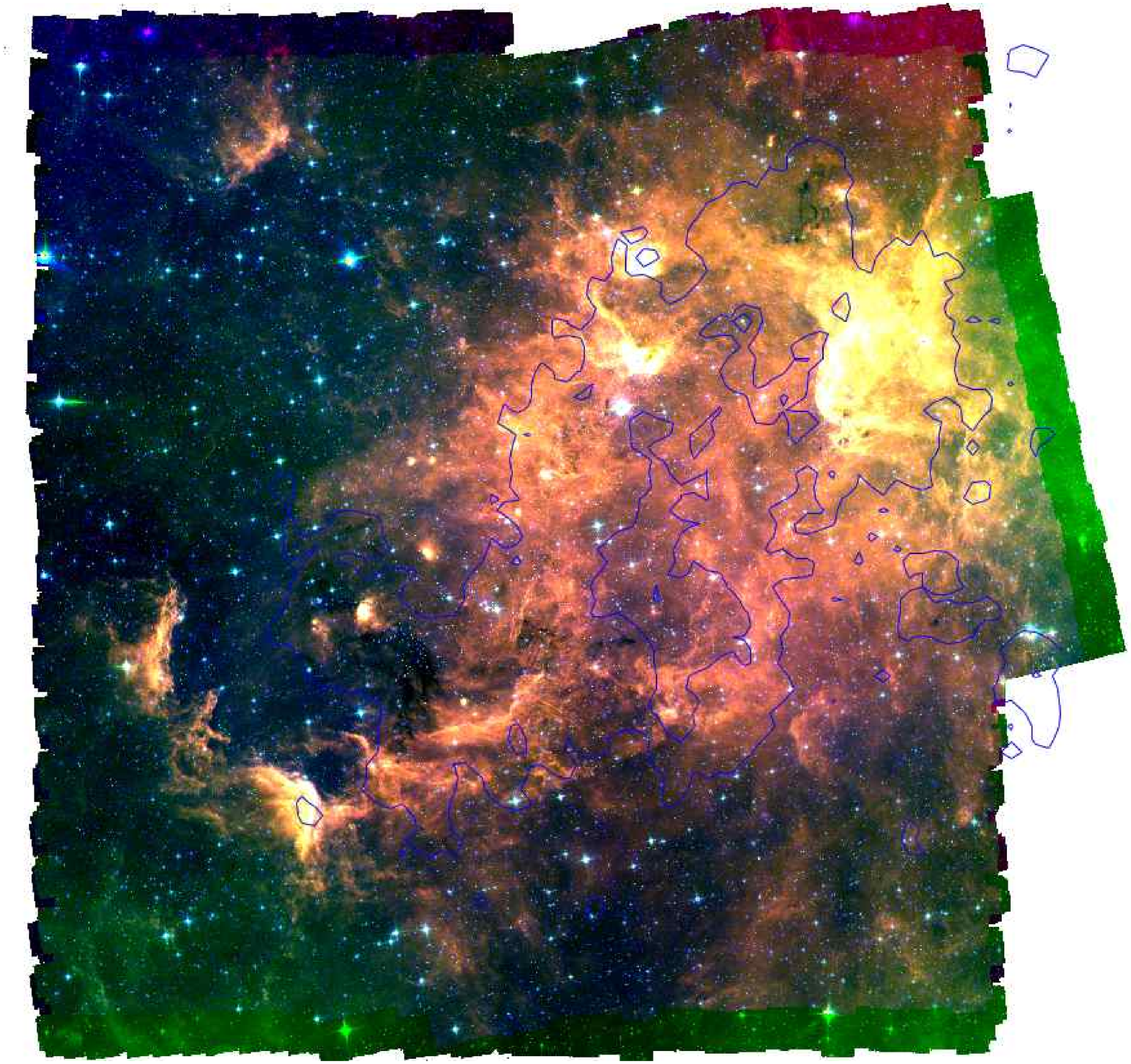}
\caption{ A 3-color view of the IRAC NANeb with 4.5~$\mu$m in blue,
5.8~$\mu$m in green, and 8~$\mu$m in red. Orientation and map center
are the same as Figure~\ref{fig:mosaic_optical}; overlays from Figure~\ref{fig:mosaic_optical} are included for reference.}  
\label{fig:mosaic_3colors}
\end{figure*}

We started with the Spitzer Science Center (SSC) pipeline-produced
basic calibrated data (BCDs), version S14.4.  We ran the IRAC Artifact
Mitigation code written by S.\ Carey and available on the SSC website.
We constructed mosaics from the corrected BCDs using the SSC
mosaicking and point-source extraction software package MOPEX
\citep{Makovoz-2005}.  The mosaics have a pixel scale of 1.22$\arcsec$ px$^{-1}$,
very close to the native pixel scale.  Figure
\ref{fig:mosaic_I1} 
shows the mosaics in channel~1 (3.6\,$\mu$m), and Figure
\ref{fig:mosaic_3colors} shows a 3-color IRAC mosaic (4.5, 5.8, and 8\,$\mu$m).

We performed aperture photometry on the combined long- and
short-exposure mosaics separately using the output of the APEX (par of
the MOPEX software) 
detection algorithm (APEX is a part of the MOPEX package) 
and the ``aper'' IDL procedure from the IDLASTRO
library.  
We used a 2 pixel radius aperture and a sky annulus of 2-6 pixels.  The (multiplicative)
aperture corrections we used follow the values given in the IRAC Data
Handbook: 1.213, 1.234, 1.379, 1.584 for IRAC channels 1, 2, 3, and 4,
respectively. 
  Fluxes   have  been converted to magnitude given the zero
magnitude fluxes of 280.9\,$\pm$4.1, 179.7\,$\pm$2.6, 115.0\,$\pm$1.7
and 64.1\,$\pm$0.9 for channels 1, 2, 3, and 4 respectively (IRAC Data
Handbook).
We take our photometry errors to be that given by the IDL
procedure.  We have checked that the uncertainties derived
by the IDL procedure are consistent with the dispersion in IRAC colors for bright,
off-cloud sources.
For the longer integration times, the  estimated
average errors on our photometry in the 4 IRAC channels are 0.020,
0.025, 0.031, 0.035 magnitudes for bright stars, increasing to $\sim$0.1
mag for channels 1-4 at 15, 14.7, 13.2, 12.7 mag.

Since the sources have been extracted individually from the mosaiced BCDs, 
we have used the overlap between BCDs to estimate our astrometric
precision. The histogram of coordinate differences shows a one sigma
RMS uncertainty of 0.3$\arcsec$ ($\sim 1/4$ of pixel) in both directions.

The APEX source detection algorithm has a tendency to 
identify multiple sources within the PSF of a single bright source
which can cause significant confusion at the
bandmerging stage. 
Because our nominal fluxes are from
the 2 pixel aperture photometry, any object which has a companion
within 2 pixels will have a confused flux. 
The photometry lists were therefore cleaned of multiple sources prior to the bandmerging. 
Objects with a companion within 2 pixels had one of the sources removed. The 
source  chosen for removal
was the one with the lowest signal-to-noise ratio.

We extracted photometry from the long and short exposure mosaics
separately for each channel, and merged these source lists together by
position using a search radius of 2 pixels (2.44$\arcsec$) to obtain a
catalog for each channel.  
The magnitude cutoff where we transition
to using the long rather than the short exposure photometry corresponds
to magnitudes of  11, 10, 8.4, and 7.5 magnitudes for channels
1, 2, 3 and 4 respectively.

Because of the complex nebulosity in this region, APEX (like any
point-source detection algorithm) can be fooled by structure in the
nebulosity.  This effect is most apparent at 8\,$\mu$m, where there are
also the fewest stellar (point) sources detected. 
To attempt to remove
false sources, we have compared photometry obtained via a 2 and
3-pixel aperture. After applying the aperture correction given in the
IRAC Data Handbook, we eliminated sources with a difference of
magnitude $>0.2$ mag; doing so rejected 32\% of the raw detections
in  IRAC's 8\,$\mu$m band.  This process is only applied to the 8\,$\mu$m channel
because the nebulosity is most prominent there (due to the presence of strong
PAH bands at 7.6 and 8.6\,$\mu$m). Also, because there are
the fewest point sources detected in this band (see Table~\ref{tab:stat}), the chances of there
being two legitimate sources within 2-3 pixels of each other are much
lower than in channels 1 or 2.  Spot-checking the images
confirms this assertion. We acknowledge this step may eliminate some
true YSOs from our 4-band catalog. We deem this to be acceptable
because our goal is to produce a minimally contaminated catalog of YSOs, not
a complete catalog. 

\subsection{Ancillary Data \& Bandmerging}
\label{sec:bandmerging}
Table \ref{tab:stat} provides statistics for our entire catalog of objects;
an object is included in our master catalog even if it is only detected
in a single band.   In order to create our final multi-wavelength catalog
which we will use to identify new YSOs,
we first merged the four individual IRAC source lists together,
starting with IRAC-1, taking the closest source within 1$\arcsec$ as
the best match. The radius of 1$\arcsec$ has been chosen based on
our astrometric precision of 0.3$\arcsec$ (see section above) and the
density of sources in
the 3.6\,$\mu$m image ({\it i.e.} the most crowded image).

In order to provide photometry at other bands, we cross-matched
our catalog to 2MASS, taking the closest source within 1$\arcsec$, and
then to our optical catalog, again taking the closest source within
1$\arcsec$.  Out of the 63\,084 objects
detected at all 4 IRAC channels, 93\%  have a 2MASS
counterpart. For  the entire region, 10\% of the final catalog stars have
an optical
counterpart; out of just the region covered by the optical photometry
data, 28\% have an optical counterpart.
The MIPS catalog will be discussed in detail in R09. In summary, we have covered approximately the same area as the 
IRAC map, and we performed the source extraction using APEX.

$BVI$ images were obtained with the KPNO~0.9\,m telescope on four photometric
nights in June 1997. An area of approximately 2.4$\times$1.7\,deg$^2$ was covered by mosaicing
the $23.2^\prime \times 23.2^\prime$ FOV CCD in a grid with overlap of typically $3^\prime$
along each border.  Exposures of 10 sec and 500 sec enabled unsaturated
photometry between $V=11$ and 20 mag.  In IRAF, images were bias subtracted
and flat fielded with sky flats taken each night.  Sources were identified
using a 5$\sigma$ detection threshold and the DAOFIND task.  The photometry
was measured through an aperture of 5 pixels radius with background determined
as the median in an annulus from 7-14 pixels where the plate scale was
0.68$^{\prime\prime}$/pixel.  Aperture corrections were applied.  Absolute photometric
calibration was achieved through observation of Landolt standards and further
self-calibrated across the overlap regions using stars in common
between frames.  First the short frames were calibrated to the long frames
then the long frames were tied across the distinct telescope pointings,
ignoring individual photometric errors larger than 0.1 mag.
The average of the 43 spatial frame-to-frame offsets was 0.024 mag
with dispersion 0.041 mag and a few frames requiring offsets as large as
0.1 mag. The individual offsets themselves have standard deviation typically
0.02-0.03 mag
and standard deviation in the mean $<$0.01 mag, which we take as
the self-calibration error.  Photometry was adopted from the long frames
except when in the nonlinear or saturated regimes of the CCD response
in which case the short frame photometry was adopted; these criteria
were applied at $B<$15.5, $V<$14.5, $I<$13.5.
Astrometry was obtained using the HST Guide Star Catalog and
the TASTROM task and is estimated accurate to $<$0.3" ($1\sigma$)
based on stars in the overlap regions.

Additional $RI$ frames were obtained on a fifth night over a more
limited area, $2\times 1.5\,$deg$^2$.  These were reduced in a similar way, requiring
a 5$\sigma$ detection threshold.

\begin{deluxetable*}{lrcccccccc}
\tablecaption{\update{03/09}
Numbers of IRAC sources extracted from the NANeb data.
\label{tab:stat}}
\tabletypesize{\scriptsize}
\tablehead{%
\colhead{Band} & \colhead{Number of sources}&  \multicolumn{7}{c}{Fractional number
  of sources which match (in \%)\tablenotemark{a}}\\ & & \colhead{$[3.6]$} & \colhead{$[4.5]$} &
\colhead{$[5.8]$} & \colhead{$[8.0]$} & \colhead{$[3.6+4.5]$} & \colhead{$[3.6+4.5+5.8]$} &
\colhead{$[3.6+4.5+5.8+8.0]$}%
}
\startdata
 $[3.6]$ & 558053 & \ldots&    75 &    32 &    12 &    75 &       30 &          11\\
 $[4.5]$ & 510922 &    82 & \ldots&    33 &    13 &    82 &       33 &          12\\
 $[5.8]$ & 200875 &    90 &    83 & \ldots&    32 &    83 &       83 &          31\\
 $[8.0]$ &  77898 &    84 &    83 &    82 & \ldots&    83 &       81 &          81\\
 $JHK_s$  & 220799 &    90 &    86 &    65 &    27 &    86 &      64 &          26\\
$BVI$ &  16595 &    85 &    83 &    72 &    40 &    83 &       71 &          39\\
 $[24]$ &   4232 &    51 &    51 &    48 &    43 &    51 &       48 &          43\\
\enddata
\tablenotetext{a}{As an example of how to interpret this table, the ``75" in cell ([3.6], [4.5]) means that 75\% of the stars detected at 3.6\,$\mu$m also are detected at 4.5\,$\mu$m}
\end{deluxetable*}

\section{Identification and Characterization of our YSO Sample}
\label{sec:ysos}

\subsection{Selection of YSO candidates}

Several studies in the literature, usually focusing on subregions of
this complex, have identified a total of $\sim$170 YSOs in the area
covered by our IRAC map, using a variety of techniques such as NIR
excess or H$\alpha$ emission.  Most of these objects are earlier than K7. 
A complete list of the new
Spitzer data for these previously-known YSOs will appear in R09.

Now, with our new, comprehensive multi-wavelength view of the complex,
we can begin to create a census of 
the members of this complex having infrared excesses. 
However, doing so is difficult, as it requires an
extensive weeding-out of the galactic and extragalactic contaminants.
To begin this task, we have opted to create a minimally contaminated sample of Spitzer-selected YSO 
candidates, as opposed to identifying every possible YSO
candidate. 
By miniminally contaminated sample, we simply mean a sample which includes as
few non-YSOs as possible ({\it i.e.} from which AGB stars, AGN, and any other
objects whose IR colors mimic YSOs have been eliminated).
We identify only stars with excesses at IRAC wavelengths as YSOs -- hence
any member without excess ({\it i.e.} Class~III YSOs) or stars with
excess only at wavelengths greater than 8\,$\mu$m (inner cleared
regions or transitional disks) will not be included.
We now discuss the selection criteria we have used to select our reliable YSO sample.

\begin{figure*}\gpos
\plottwo{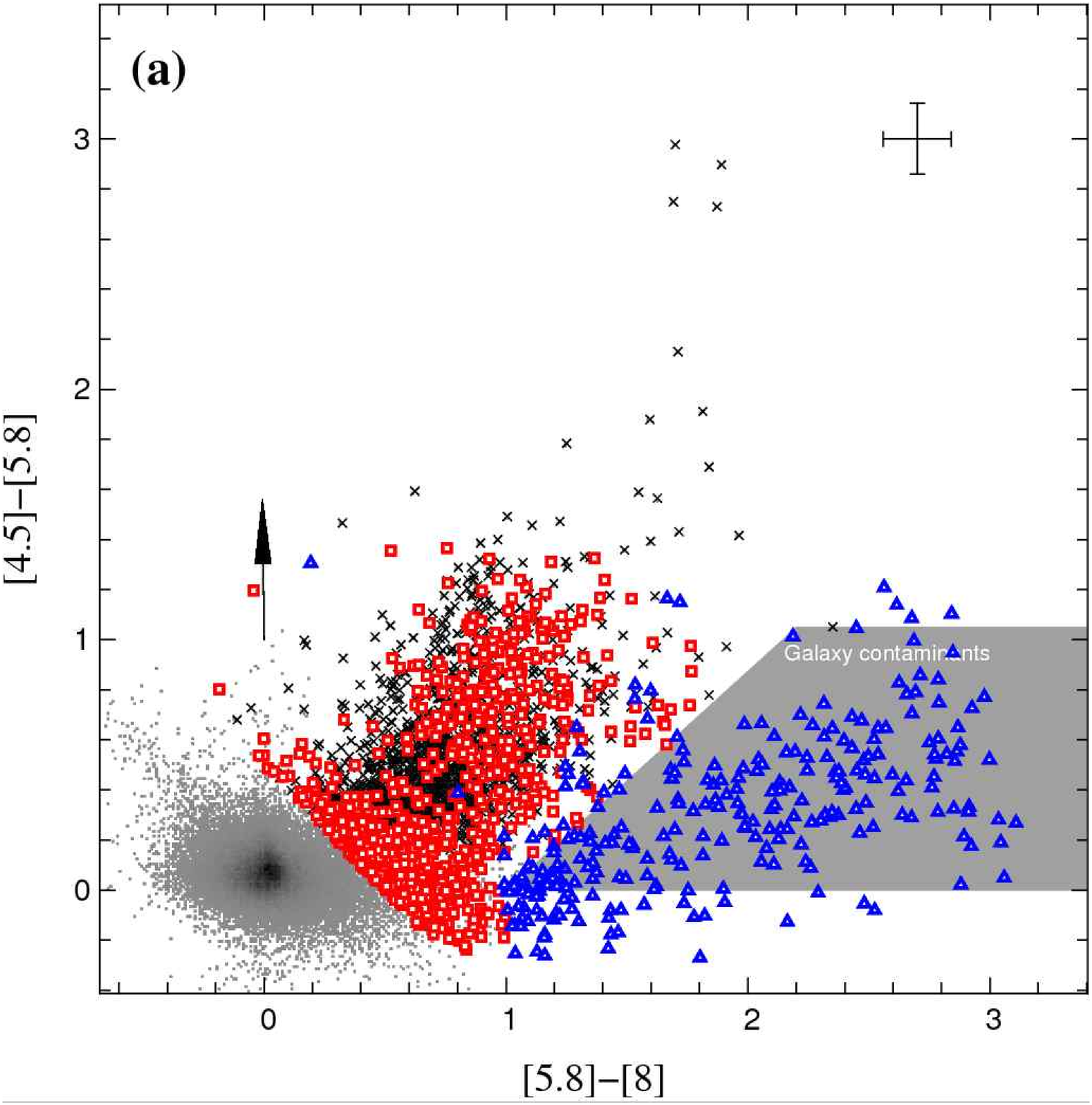}{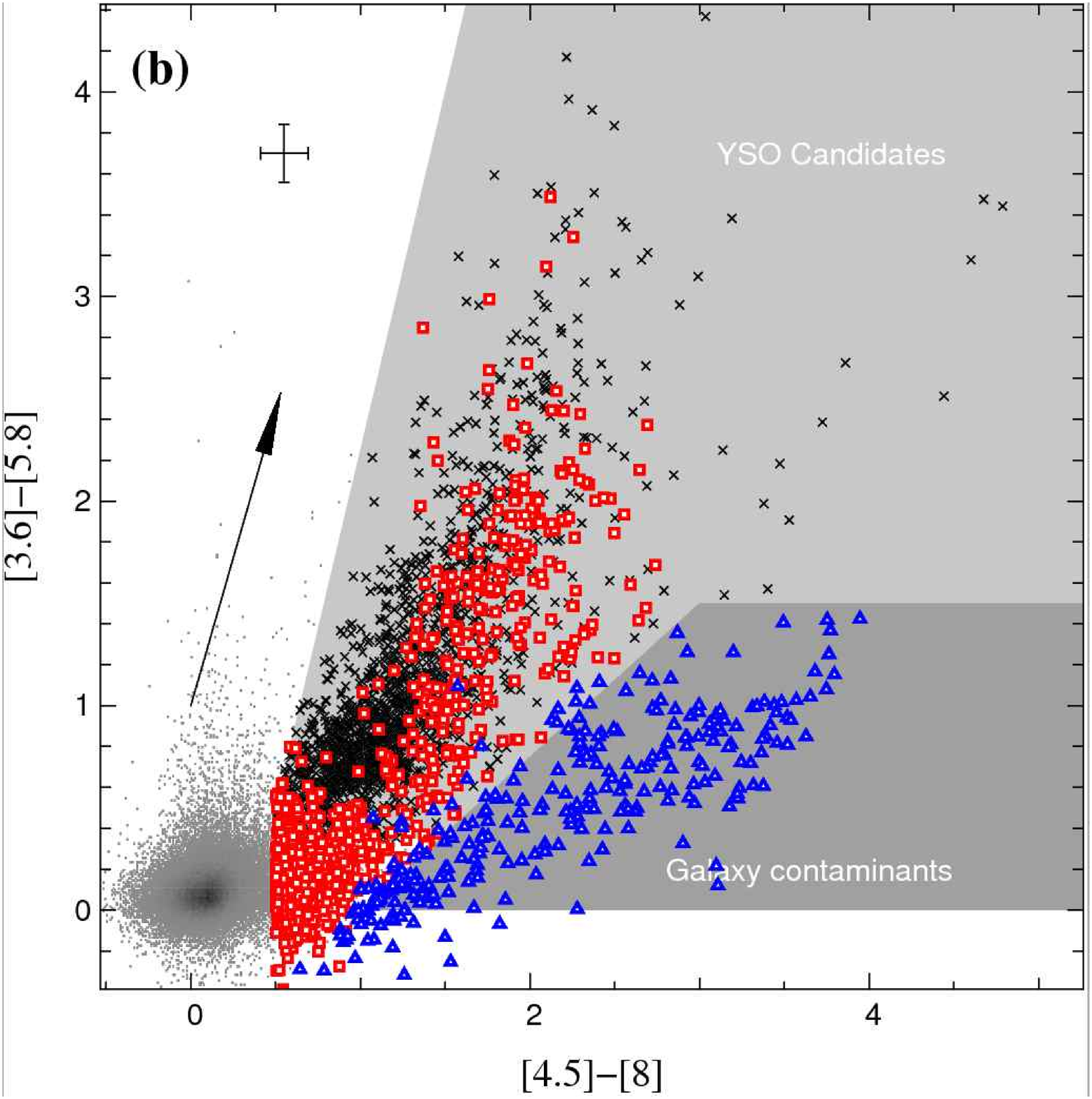}
\plottwo{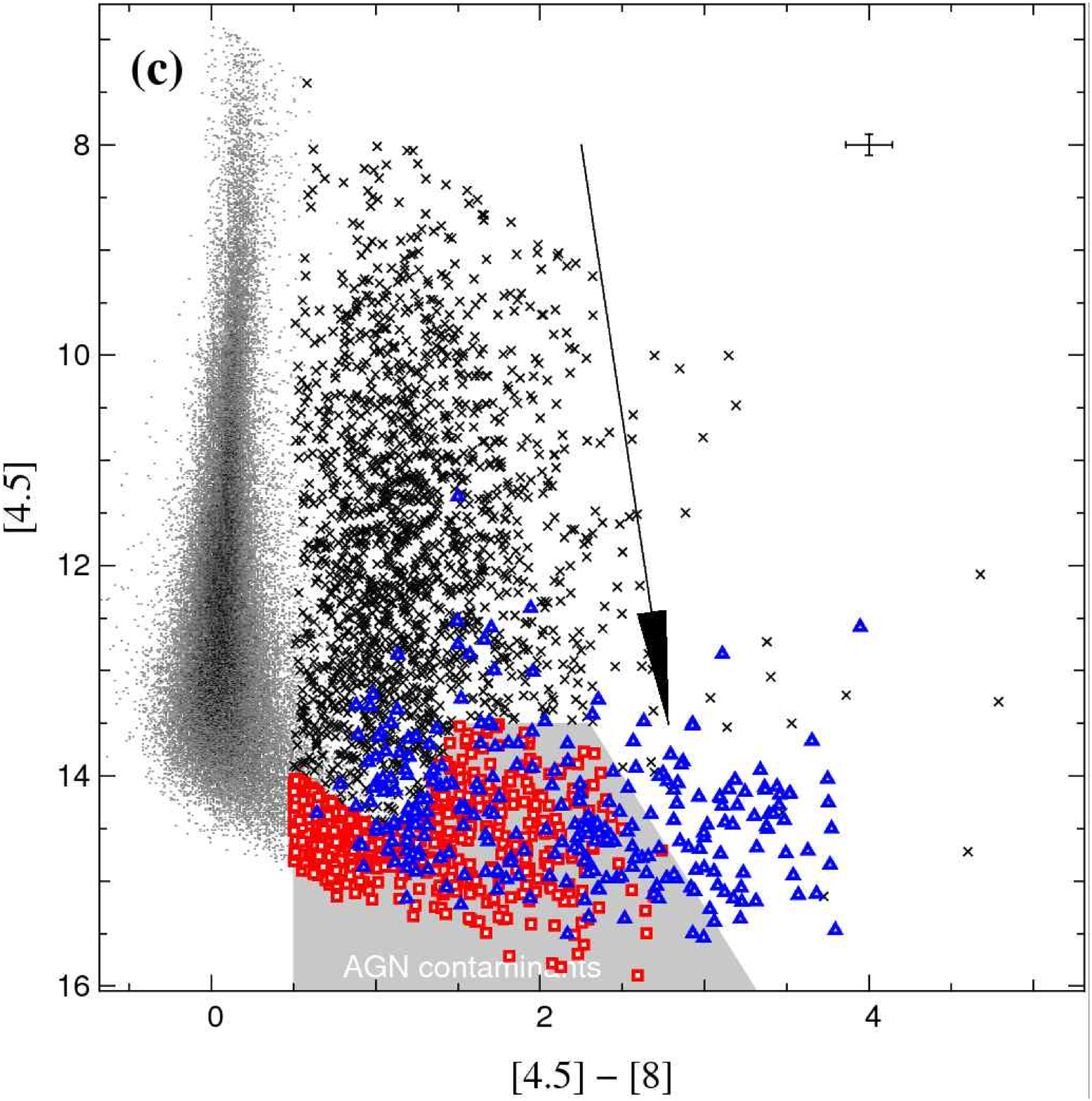}{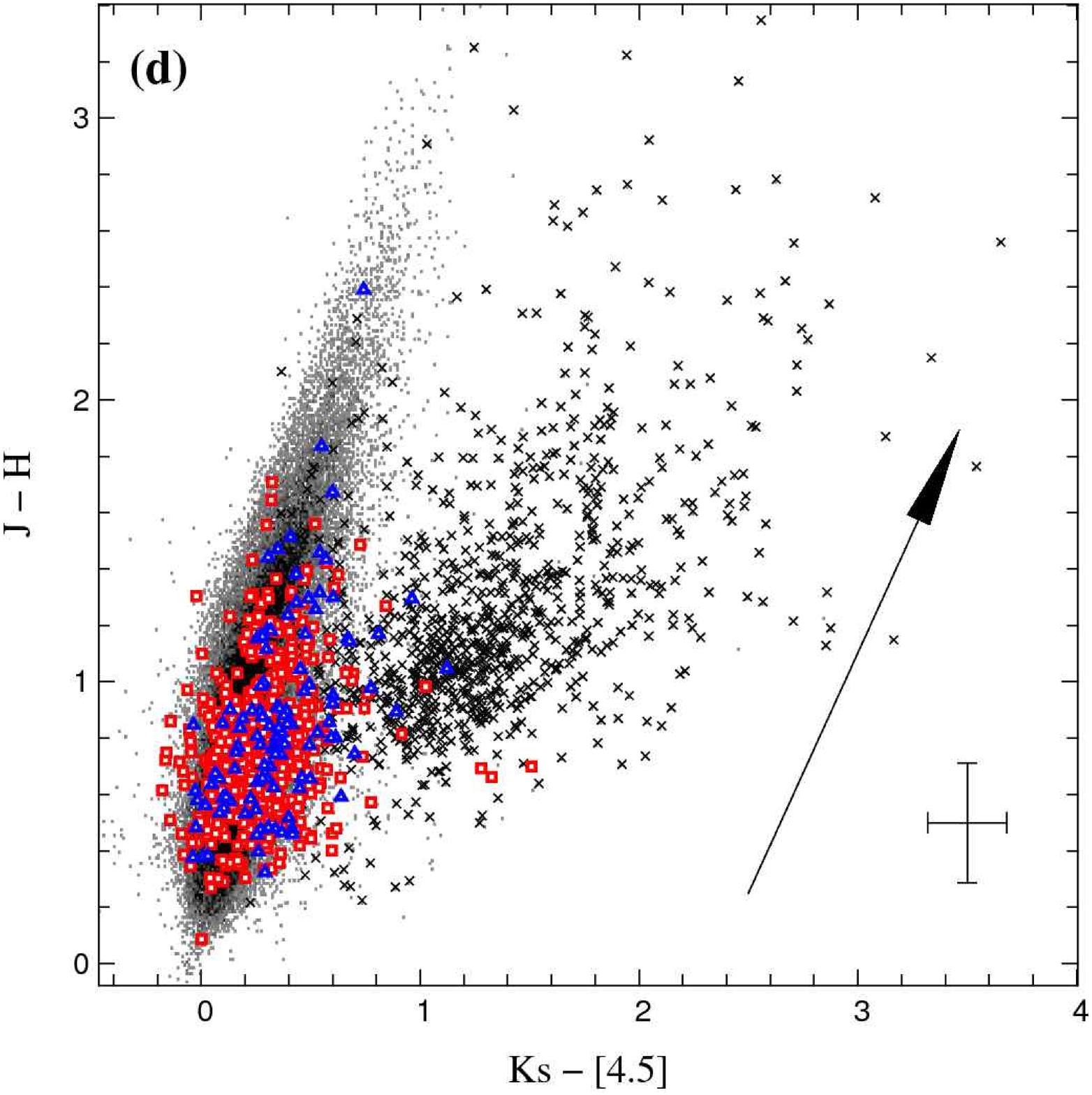}
\caption{IRAC color-color and color magnitude diagrams used to reject
  background contaminants (panel {\it a}, {\it b} and {\it c}) and select YSOs
  candidates (panel {\it b}). An additional diagram shows the distribution
  of background contaminants and YSO candidates  in a
  2MASS-IRAC color-color diagram limited to targets with 2MASS
  counterparts (panel {\it d}).
  In all diagrams, blue triangles are sources flagged as PAH-emission
  sources (galaxies) located in the darkest gray area of
  panel {\it a}
  or {\it b}. Red squares are sources flagged as AGN, selected in the gray
  area of panel {\it c}. Finally, YSOs candidates are plotted with black cross
  symbols; they have been selected in the brightest gray area of panel
  {\it b} if not previously flagged as background contaminants. 
  The reddening vectors correspond to an $A_K$ of 10 in panels {\it a},
  {\it b} and {\it c}, and $A_K$ of 2 in panel {\it d}. We averaged the
  extinction law given in \cite{Flaherty-2007} for Serpens, Orion and Ophiuchus.}
\label{fig:iraccolorcolor}
\end{figure*}

Our selection method is described below and illustrated in Figure \ref{fig:iraccolorcolor}. 
We initially require detection at all four IRAC bands, which substantially
limits the catalog (see Table~\ref{tab:stat}) and then we follow
the IRAC four-band source  characterization described in
\citet[][section~4.1]{Gutermuth-2008a}.   
Using their extragalactic contamination criteria, we have first rejected 272
sources as having colors consistent with galaxies dominated by PAH
emission. This selection has been made in the $[4.5]-[5.8] / [5.8]-[8]$
and $[3.6]-[5.8] / [4.5]-[5.8]$ planes, where we reject objects seen
in the grey zones plotted in panel (a) and (b)
of Figure \ref{fig:iraccolorcolor}; this selection is defined by two
sets of inclusive equations: 
\begin{equation}
\begin{array}{c}
\left\{
\begin{array}{c}
    [4.5] - [5.8] < \frac{1.05}{1.2}([5.8]-[8.0]-1) \\
    \& \\
    \left[4.5\right]-[5.8] <  1.05 \\
    \& \\
    \left[5.8\right]-[8.0] > 1    
\end{array}\right.\\\\
\ \ \mathrm{OR} \ \ 
\\\\
\left\{
\begin{array}{c}
    [3.6]-[5.8]<\frac{1.5}{2} ([4.5]-[8.0]-1) \\
    \& \\
    \left[3.6\right]-[5.8] <  1.5 \\
    \& \\
    \left[4.5\right]-[8.0] > 1    
\end{array}\right.
\end{array}
\end{equation}

We then rejected 823 more  sources having colors consistent
with AGN in the [4.5] vs.\ [4.5]$-$[8.0] plane; the region of color
space used to
select AGN-like sources is plotted in panel (c) of Figure
\ref{fig:iraccolorcolor} and is defined by the following
equations: 
\begin{equation}
\begin{array}{c}
\left\{
\begin{array}{c}
[4.5]-[8.0] > 0.5 \\
\& \\
\left[4.5\right] > 13.5 + ([4.5]-[8.0]-2.3)/0.4 \\
\& \\
\left[4.5\right] > 13.5
\end{array}\right.\\ \\
 \ \ \&\ \ 
\\ \\
\left\{
\begin{array}{c}
[4.5] > 14 + ([4.5] -[8.0] -0.5)\\
\mathrm{OR} \\
\left[4.5\right] > 14.5 - ([4.5] - [8.0] - 1.2)/0.3 \\
\mathrm{OR} \\
\left[4.5\right] > 14.5 
\end{array}\right.
\end{array}
\end{equation} 
(see Appendix~A of \citealt{Gutermuth-2008a} for more details).  
Note that the  AGN-like source selection has been made in the observed [4.5]
vs.\ [4.5]$-$[8.0] plane, whereas \cite{Gutermuth-2008a} use the
dereddened plane.  \cite{Gutermuth-2008a} are able to work easily in the
dereddened plane because they have a high 
spatial resolution $A_v$ map. Because the NAN complex is much further
away than NGC\,1333, we do not have the ability to construct such a high-resolution $A_V$ map. 
We tested the implications of this limitation by using the
\cite{Cambresy-2002} extinction map  to deredden individual objects. 
Just 25 sources would be dropped by performing this
selection in the dereddened plane; they are fairly uniformly spread 
over the entire mapped region (and hence are likely contaminants
rather than YSOs). Just 3 of
these are in the Gulf of Mexico region (see R09) where we do not
expect very much background contamination due to the very high
reddening. Satisfied that this decision does not significantly affect
our final sample of YSOs, we have chosen to keep our source selection in the
observed plane. 

Once these background contaminants are removed from consideration, we
are left with a population of 61,989 IRAC sources, dominated by 
objects with the apparent colors of stellar 
photospheres. As in \cite{Gutermuth-2008a}, we have further selected YSO candidates in
the $[4.5]-[8]$ {\it vs.} $[3.6]-[5.8]$ color-color diagram meeting the
following criteria: 
\begin{equation}
\left\{\begin{array}{c}
\left[4.5\right]-[8]>0.5\\
\&\\
\left[3.6\right]-[5.8]>0.35\\
\&\\
\left[3.6\right]-[5.8]\le \frac{0.14}{0.04} \times \left([4.5]-[8]-
  0.5\right)+0.5 
\end{array}\right.
\end{equation}
This selection is indicated in Figure~\ref{fig:iraccolorcolor}, panel
{\it b}.
This leaves \nbyso\ candidates.\update{03/11}

Continuing to follow \cite{Gutermuth-2008a}, we used their 
selection criteria to identify objects from the \nbyso\ candidates
whose IRAC fluxes may be contaminated by emission lines from
shocks given their $[3.6]-[4.5]$ and 
$[4.5]-[5.8]$ colors by the equations:

\begin{equation}
\left\{
\begin{array}{c}
[3.6] - [4.5] > \frac{1.2}{0.55} ([4.5]-[5.8] -0.3)+0.8\\
\&\\
\left[4.5\right]-[5.8] \leq 0.85\\
\&\\
\left[3.6\right]-[4.5] > 1.05
\end{array}\right.
\label{eq:shoked}
\end{equation} 

We found only 2 YSO candidates that matched these criteria
(205608.3+433654.2  and  205702.1+433431.3). Their SEDs are 
compatible with their being real YSOs whose aperture photometry
is affected by emission from a compact, circumstellar nebula, 
causing an excess at 4.5\,$\mu$m relative to the two adjoining
IRAC bands.  Moreover, these 2 objects are located inside the ``Gulf of
Mexico'' where the concentration of YSOs is the highest in the cloud, 
and where previous optical emission line surveys have found numerous
HH regions.
Given these considerations, we retain these objects in our YSO candidate list.

Of the \nbyso\ YSO candidates, 972 (59\%) have a 2MASS counterpart\update{03/11}. Out of
the entire region, 131 (7.9\%) have an optical counterpart; out of the
region covered by the optical data, 10.4\% have an optical
counterpart. All but 63 of the YSO candidates are inside the region covered by
MIPS data\update{03/11 with MIPS patches}, and 45\% of the candidates have a MIPS 24\,$\mu$m counterpart. 
We note that the percentage of 2MASS and optical YSO counterparts is
less than for the whole IRAC 4-band catalog (92\% and 28\%
respectively, see section~\ref{sec:bandmerging}); this is due to the
fact that YSOs are mostly located inside highly reddened regions and
are not detected in our shorter wavelengths.
With this selection of $\sim$1600 YSOs, we have increased the number of likely
YSOs in the NANeb region by an order of magnitude.

We compare the positions of both the YSO candidates and
background contaminants in a 2MASS / IRAC color color diagram in 
panel {\it d} of Figure \ref{fig:iraccolorcolor}. 
A large fraction (90\%) of sources flagged as
galaxies or AGN fall in the region occupied by reddened main
sequence stars,  and hence blueward in $K_s-[4.5]$ color from
the YSOs. 
Just 5\% of the YSO candidates are located in this same region,
two-thirds of which are classified as intermediate between Class~II
and Class~III (see \S\ref{sec:classes}).   

It is hard to estimate the contamination fraction of our YSO
candidates without additional data (e.g. spectroscopic or X-ray confirmation).
However, we can make a few worst case estimates.  Because our survey area
is fairly large, it includes some sections which are outside the main
star-forming region and have relatively few candidate YSOs.  The north-east
corner of our survey region (see Figure \ref{fig:mosaic_3colors}), for example,
offers one such relatively YSO-poor area.   Specifically, we define a
``field" region as
being the area above a line from the mid-point of the East edge of
Figure \ref{fig:mosaic_3colors} to the mid-point of the North edge of
the figure.  If all the candidate YSOs in this region are contaminants,
we derive a contaminant surface density of 32 objects per square degree.
Assuming the AGN and other contaminants are uniformly distributed over our
survey region, this yields an upper limit to the fraction of contaminants
in our survey of 17\%.
  
\subsection{NANeb age estimation}
\begin{figure}
\centering
\plotone{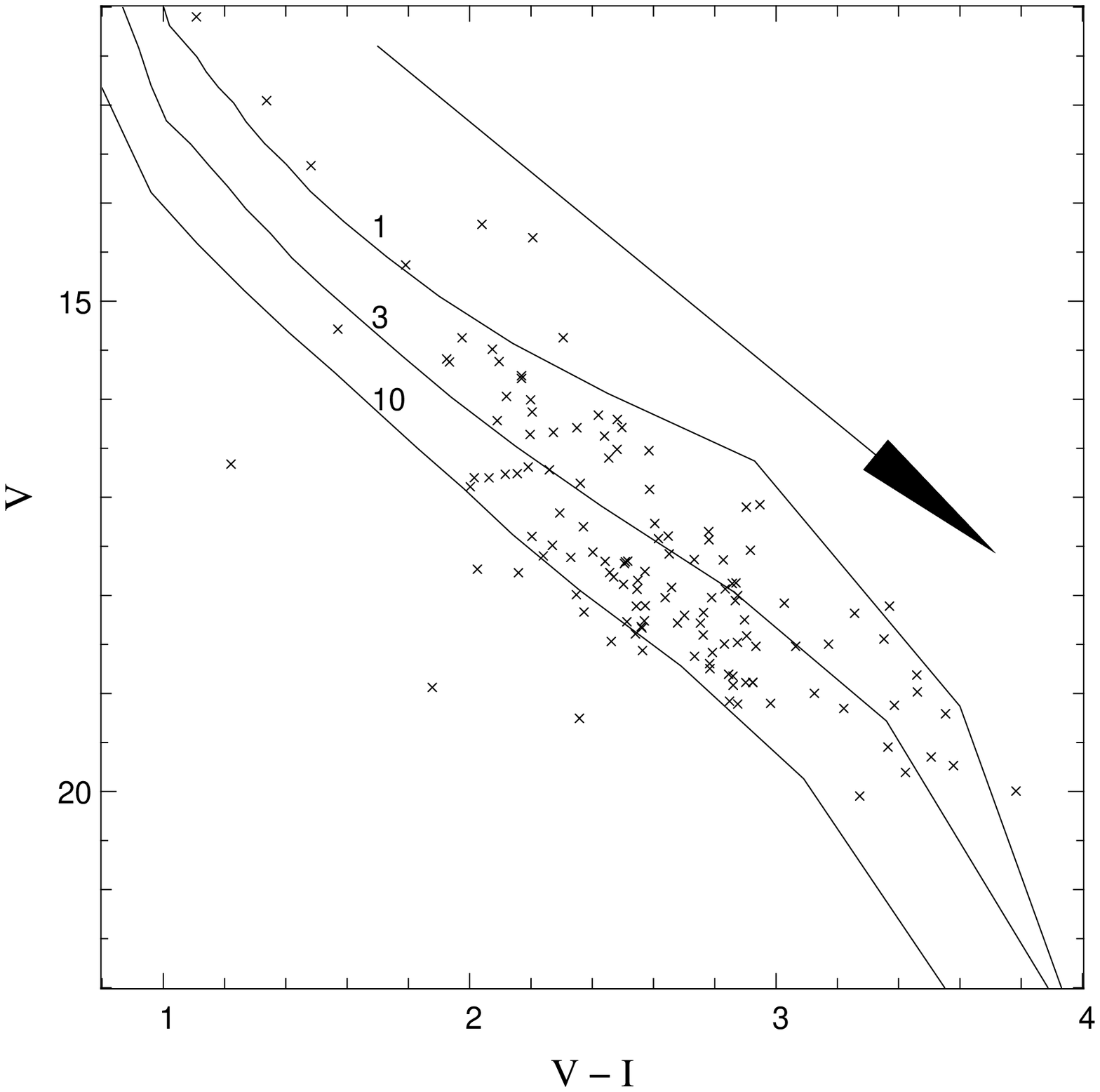}
\caption{$V/V-I$ color magnitude diagram for our YSO candidates. 
The solid lines are, from
right to left, tuned Siess models (see text) for 1 3 and 10\,Myr scaled
to the NANeb distance. A reddening vector of $A_v$=5 is plotted.}
\label{fig:vvi}
\end{figure}

We have used optical photometry in order to constrain
the age of the NANeb complex.  
Figure \ref{fig:vvi} shows the location of YSOs with optical and
2MASS photometry in an optical color-magnitude diagram. Also shown are   
\cite{Siess-2000} isochrones, where the color-$T_{\rm eff}$
conversion has been tuned so that the 100\,Myr isochrone follows
the single-star sequence in the Pleiades \citep{Stauffer-1996,Jeffries-2007}. 
The number of YSOs in this diagram is limited by the fact that it
requires optical counterparts. This means that most of sources
from this diagram are located inside the less-extincted  regions of the NANeb,
on the edge of the Pelican Nebulae (80\% are located inside region of $A_v<5$,
according to the \citealp{Cambresy-2002} extinction map). The embedded Class~I sources are not
well-detected in the optical data, and our YSO selection criteria do
not allow us to detect Class~III sources, so this sample is mostly
limited to  Class~II sources.  

Our age estimates are uncertain for all of the normally expected
reasons (e.g. uncertainties in the isochrones and their transfer to
the observational plane; the need for and imprecision of the reddening
corrections; the effect of spots and UV excesses on the estimated
colors and effective temperatures of our target stars; binarity).
However, it is apparent in Figure
\ref{fig:vvi} that most of our YSO candidates (with optical 
counterparts) are younger than 10~Myr according to the isochrone
tracks. The median age is slightly older than 3\,Myr, but 
the most embedded and probably youngest regions of NANeb are excluded
from this optical CMD. The median age and the age dispersion is
comparable to what found in NGC~2264 by \cite{Rebull-2002}. 
Since the reddening vector is essentially parallel to the isochrones,
and YSOs with optical counterpart have low reddening,  
the median age will be minimally affected by reddening.

\subsection{Classes of YSO Candidates}
\label{sec:classes}

\begin{figure}\gpos 
\plotone{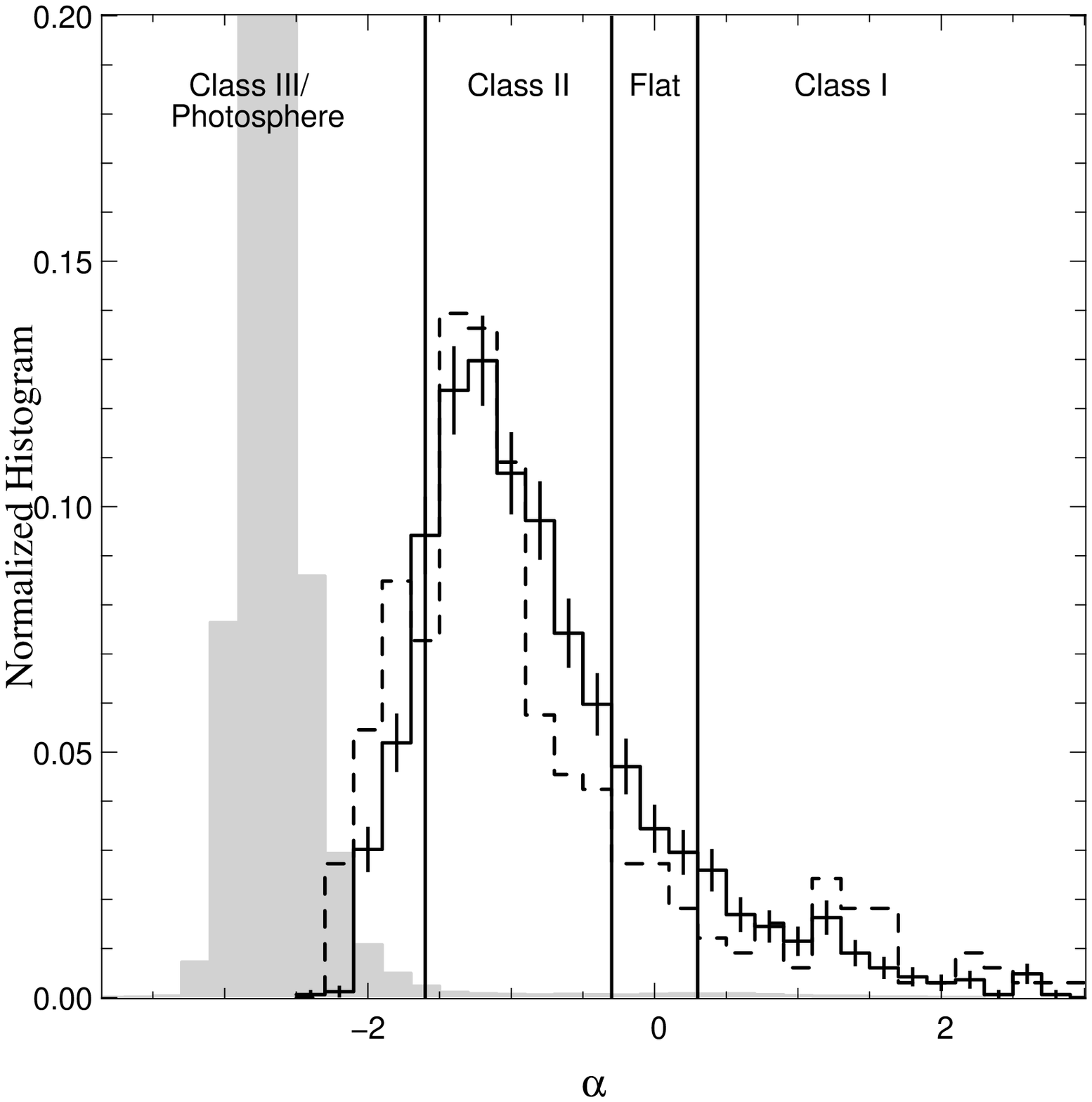}
\caption{ Normalized histograms of the distribution of spectral slope,
$\alpha$. The continuous line is the histogram for the YSO candidates
for the whole sample. All other IRAC sources from the NANeb are also
indicated (in gray), as are the 330 YSO candidates we have found from the
\cite{Harvey-2006} highly reliable catalog (dashed line).}
\label{fig:alphahisto}
\end{figure}

We classified this 
set of highly likely NANeb YSOs, 
according to the \citet{Lada-1987} Class I/II/III system, updated by
\cite{Andre-1994} where the ``flat'' class has been added between
Class~I and II.   Classes are assigned based on the spectral index,  
defined by  $$ \alpha = \frac{d\ \log{ \lambda
F(\lambda)}}{d\,\log{\lambda}} $$  where $F(\lambda)$ is the flux at
the wavelength $\lambda$. For Class~I, 0.3$\le\alpha$; for flat
sources, $-0.3\le \alpha < 0.3$; for Class\,II, $-1.6\le \alpha <
-0.3$; and for Class~III, $\alpha<-1.6$. Note that because we are
making a Spitzer-based selection of YSO candidates, by definition {\em
none} of our candidates will have a photospheric slope, so our
Class~III inventory is guaranteed to be incomplete. Further
observations at other wavelengths ({\it e.g.} with Chandra) are
required to find such objects. 
In the following, to avoid confusion, we call our YSO candidates with $\alpha<-1.6$
Class~\III.    
\sly{Cite peoples who use IRAC only to define Classes}

We fit  $\alpha$ over the 4 IRAC channel points from 3.6 to 8
$\mu$m.\update{03/11} From the total \nbyso\
YSO candidates, we found that 1059 (64\%) are Class~II, 184 (11\%) are flat and
198 (12\%) are Class~I. We classified the remaining 216 (13\%) YSOs as
Class~\III\ because they have $\alpha$ between stellar photospheres
($\alpha = -2.7$) and Class~II ($\alpha = -1.6$).   A histogram of $\alpha$
for our YSO candidate sample is plotted in Figure~\ref{fig:alphahisto}.
This figure also contains, for comparison, a histogram
of the values of $\alpha$  for our entire NANeb catalog (mostly field
stars),
and that for the YSO candidates derived from the highly reliable Serpens catalog
obtained by \cite{Harvey-2006,Harvey-2007a,Harvey-2007b} (see
discussion below).

Plots similar to those appearing
here (or in R09) can be found in most of the c2d series of papers on
Serpens, Perseus, Ophiuchus, Chamaeleon II, and Lupus.
We picked Serpens for our primary comparison here 
because it is thought to be similar
in age to the NANeb, it contains a deeply embedded
cluster similar to the Gulf of Mexico, it
is in  the Galactic plane ($l$=32, $b$=+5), and
\cite{Harvey-2006,Harvey-2007a,Harvey-2007b} performed a
Spitzer-specific source selection. However, there 
are some important differences. The Spitzer maps of Serpens are only
0.85\,deg$^2$, nearly 6 times smaller than our map.  The selection
method for finding YSO candidates is much different in
\cite{Harvey-2006,Harvey-2007a,Harvey-2007b}, for example, requiring a
MIPS-24 detection, and the differences are most strongly apparent in
the selection of Class~III objects.  

In order to fairly compare our data with the 
Serpens data, we used the highly reliable Serpens catalog (of all
objects, not just their YSO candidates) provided by
the c2d team as part of their final Legacy delivery (available on the
SSC website) and applied our selection method to those data. From the sample of
Serpens YSOs which pass our selection criteria, we calculated $\alpha$ in the
same way that we did for the NANeb, from 3.6 to 8\,$\mu$m.  It is these
values which appear in Figure~\ref{fig:alphahisto}.
Our selection and classification scheme yield 52 Class~I, 24 Flat
disk, 189 Class~II and 65 Class~\III\ Serpens YSOs. The
number of Class~\III\ objects is hard to compare to that found in
\cite{Harvey-2006} since they use the 24$\mu$m data in their
classification scheme ; but the number of Class~I, Flat and Class~II
are roughly comparable: 30, 33, 163 respectively in \cite{Harvey-2006}. 

As a means to estimate a comparative evolutionary age for
the NANeb stars, one can determine the ratio of the number
of Class\,II (``older'') to Class\,I or Flat sources (``younger''), and
compare that ratio to that derived for other clusters, such
as Serpens.   For our entire NANeb YSO sample, we derive
a ratio N(Class\,II)/N(Class\,I+Flat)  =  1059/382 = $2.78\pm0.17$.
Using our similarly analysed data for Serpens, we derive
this ratio as $2.49\pm0.33$ (roughly comparable to what
was derived by Harvey et al. 2006 - $2.6\pm0.4$, where they
used 2MASS through 24\,$\mu$m data to compute alpha).  The ratios
for NANeb and Serpens are comparable, suggesting that the
mean ages of the two YSO samples are indeed similar.

\cite{Harvey-2006,Harvey-2007a,Harvey-2007b} used a series of
Spitzer-based selection criteria to select a sample of YSOs;
their final selection in \cite{Harvey-2007b} primarily uses
color-color and color-magnitude diagrams to assign a liklihood that a
given object is extragalactic. The process starts with requiring
detection in all 4 IRAC bands as well as MIPS-24. Only about half of our
minimally contaminated YSO sample has a MIPS-24 detection, so the Harvey \etal\
selection process by its very nature could only retrieve about half of our
sample. However, $\sim$90\% of our highly reliable sample with MIPS-24
detections is also retrieved by a Harvey \etal-style source selection.

R09 contains a special discussion of the Gulf of
Mexico, which is an interesting area containing hundreds of deeply
embedded young stars, mostly associated with 3 subclusters.  We note here that our requirement for having
all four bands of IRAC for our source selection omits many of the
objects in this region, and this requirement will be relaxed
in R09 to find cluster members.  However, using our selection criteria and our best
possible YSO sample (as defined above) results in a ratio of
N(Class~II)/N(Class~I+Flat) that is statistically significantly
different inside the Gulf cluster than outside of it. The ratio, with its Poissonian error, is
901/261=3.45\,$\pm$0.24 outside the Gulf of Mexico and 
158/121=1.31\,$\pm$0.16 inside this region.  Moreover, we
find comparatively very few Class~\III\ objects in the Gulf
of Mexico region.  Both these findings indicate that the Gulf
of Mexico cluster is, in evolutionary terms, the youngest
region within the NAN complex.  
\update{03/11}

We explore now, how the NANeb reddening can change our YSO
classification. Indeed \cite{Muench-2007}, appendix~A, demonstrated that for a
deeply embedded cluster (A$_v \sim$40), the IRAC SED slope of a
typical K6 Class~II member of IC\,348 would be reddened into an apparent flat
spectrum. 
According to the \cite{Cambresy-2002} extinction map, however, 97\% of YSOs are
located in region where the A$_V$ is less than 10.  Since the
\cite{Cambresy-2002} map yields extinctions up to about 30, we assume that
for extinctions $\leq$ 10 the predictions should be reliable (i.e. not
``saturated" because the extinctions are so high that no background stars
are present within the grid point).  To test the effect of
reddening on our classification, we have dereddened our SEDs prior
to the spectral slope calculation, according to the Cambresy et
al. map. 
The numbers of Class~I, Flat disk and Class~II given by the dereddened
SEDs  - 186, 177, 1043 - are quite similar to those using the 
observed SEDs - 198, 184
and 1059, making the N(Class~II)/N(Class~I+Flat)
ratio equal to 2.87$\pm$0.17 instead of 2.78$\pm$0.17. This small 
differences does not change our main conclusions about the spatial 
distribution of Class~I and Class~II.  

There is another means by which we can gauge the effect of extinction
on our YSO classification.
\cite{Gutermuth-2008a} distinguished protostars and Class~II
by there $[4.5]-[5.8]$ color excess.   They chose this criterion
specifically because it is relatively immune to extinction (because
the interstellar extinction curve is relatively flat between these
two wavelengths).  \cite{Gutermuth-2008a}
selected protostars as
sources matching the following equation:

\begin{equation}
\left\{
\begin{array}{c}
 \left[4.5\right]- [5.8] > 1 \\
 \mathrm{OR} \\
 \left[4.5\right]-[5.8] > 0.7 \& [3.6]-[4.5]>0.7
\end{array}\right.
\label{eq:protostars}
\end{equation}

\begin{figure}\gpos 
\plotone{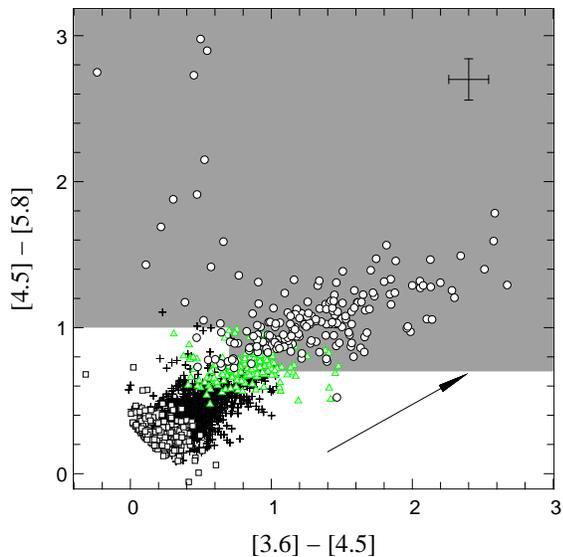}
\caption{Spitzer color-color diagram. Open squares are Class~\III, plus
  symbols are Class~II, green triangles are classified as Flat spectrum sources
  and circles are
  Class~I. The green area is the protostar selection area defined
  by \cite{Gutermuth-2008a}. A reddening vector of A$_k$=10 is plotted.}
\label{fig:protostars}
\end{figure}

The \cite{Gutermuth-2008a} classification scheme does not include
flat-spectrum sources as a separate class.
Figure \ref{fig:protostars} shows the location of Class~I / Flat /
Class~II and Class~\III\ YSOs in the  $[4.5]-[8.0] / [3.6]-[4.5]$
plan superimposed on the protostar area defined above. The location of 
our Class~I and Class~II sources agree well with the regions
defined in this diagram:
95\% of our Class~I and 99\% of our Class~II sources
would be classified as protostars and Class~II by the Gutermuth et
al. criteria. 
Our flat spectrum sources lie on the border between Class~I and
Class~II, with roughly equal numbers on either side of the
\cite{Gutermuth-2008a} boundary.

\subsection{Spatial Distribution of YSO Candidates}
\label{sec:spatial}

\begin{figure*}
\centering 
\epsscale{0.5}
\plotone{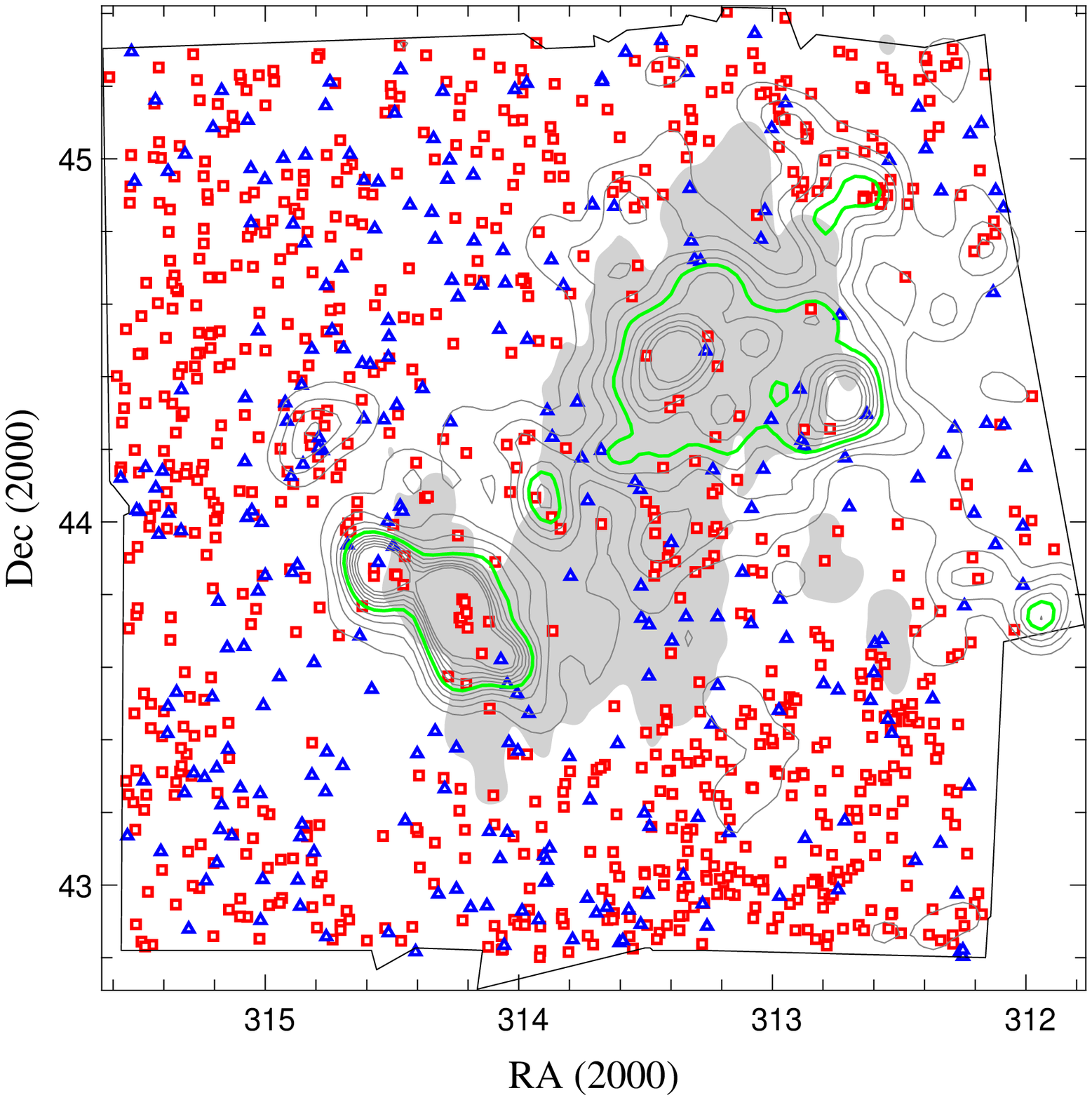}
\plotone{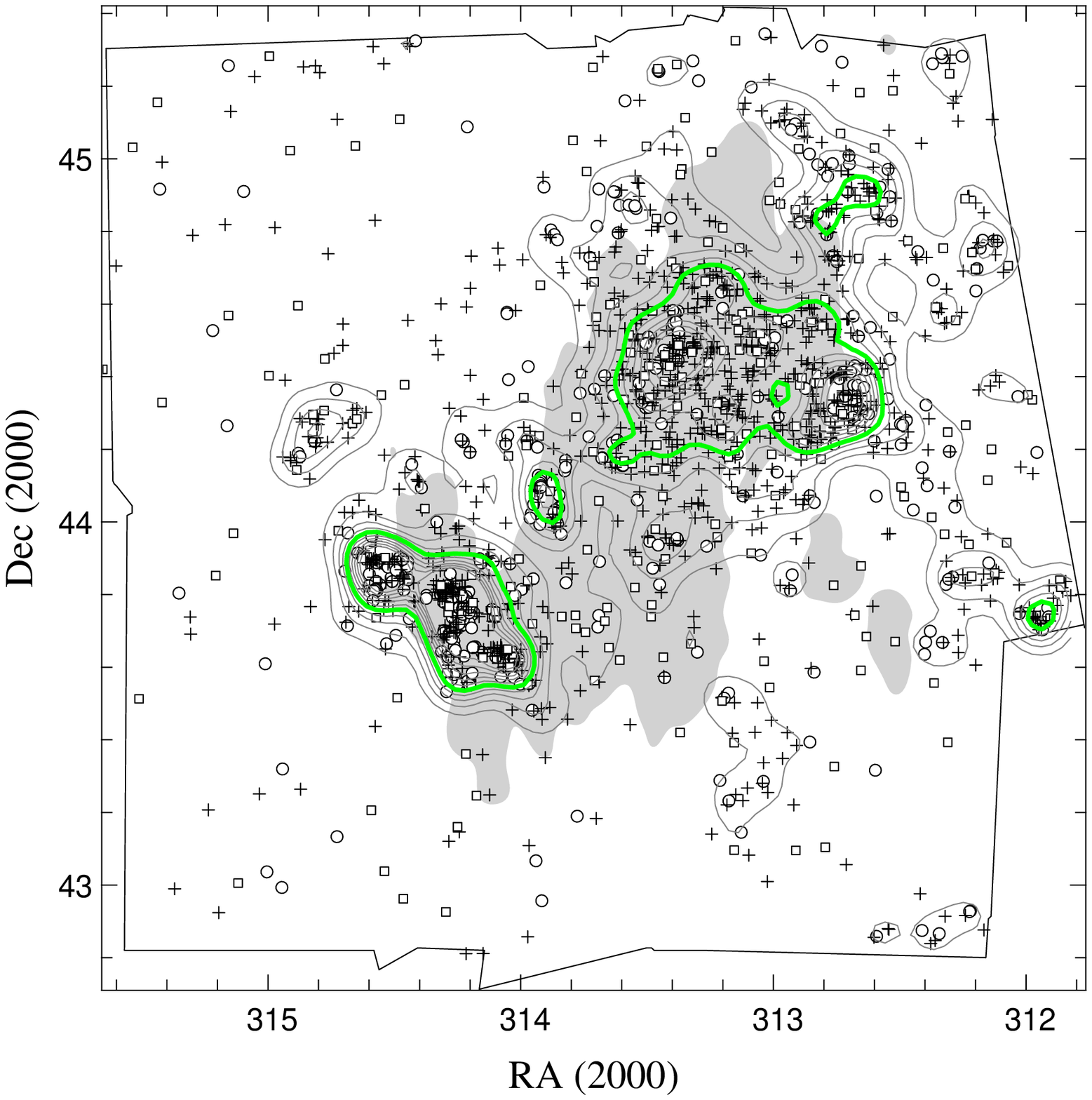}
\epsscale{1}
\caption{TOP: the spatial distribution of background
  contaminants (red squares: AGN like sources, blue triangles:
  galaxies; see text and Figure \ref{fig:iraccolorcolor}). BOTTOM: 
  the spatial distribution of YSO candidates, where symbols
  denote the object Class (circles: Class\,I or Flat, $+$:
  Class\,II, square: Class\,\III). 
  In both panels, we plot contours of YSO density (see text for more details);
  the contour levels correspond to values plotted in Figure
  \ref{fig:evolR}, from
  200 to 2000 YSOs/deg$^2$. 
  The contour which surrounds
  half of the total YSO population ($\sim$1000 YSOs/deg$^2$) is
  highlighted in green and with a thicker contour line. The background gray area
  indicates the $A_V\geq$5 dark cloud from the
  \cite{Cambresy-2002} map.}
\label{fig:distrib} 
\end{figure*}

Now that we have a minimally contaminated sample of YSO candidates, we can
look at their spatial distribution and investigate the degree of
clustering and the relative spatial distribution of YSOs as a function of
their class.  
We have computed a YSO density map of our observed region. 
We used a Kernel method \citep{Silverman-1986} which
yields a smooth isodensity contour map from the projected position of
objects. In each point of the space $(\alpha, \delta)$, the density is
set, due to the contribution of all $n$ points, by the kernel
density estimator:  $$D(\alpha, \delta) = \frac{1}{h^2}\sum_{i=1}^{n}
K(\alpha, \alpha_i, \delta, \delta_i)$$ where $K$ is the kernel. For
the kernel, we adopted a Gaussian shape :  $$K(\alpha, \alpha_i,
\delta, \delta_i) = \frac{1}{2\pi} \exp{ \frac{-\left(
(\delta-\delta_i)^2 + (\alpha-\alpha_i)^2 \cos^2\delta 
\right)}{2h^2}}$$  where $h$ is the smoothing parameter. We adopted
$h=0.05\arcdeg$ (0.5 pc), which is approximately the size of the smallest group of YSOs
in the NANeb discernible `by eye'. 

Figure \ref{fig:distrib} shows contours of the density map we
obtained, where both  background contaminants and YSO candidates  are
plotted in two different panels.  Using these density contours, we
find first that the background contaminants (galaxies or AGN) are
relatively evenly distributed across the field of view.  However, we
notice a lack of background contaminant sources on the central dark
cloud.  The lack of contaminants in the high extinction regions can be
quantitatively explained simply due to the fact that our survey is
magnitude-limited, and the extra extinction causes the observed fraction
of galaxies to drop out of our sample. 
We checked this assumption by creating an artificial map of randomly distributed
background sources reddened by the extinction map of
\cite{Cambresy-2002}.   We then applied a magnitude cut to this artificial
sample, corresponding to the effective faint limit imposed by our 4-band
YSO selection criteria.   The resulting spatial distribution of
selected galaxies mimics well the observed galaxy distribution 
in Figure 7(top), confirming our assumption. 

Our YSO candidates are located primarily on the central dark cloud
(Lynds 935), and are for the most part highly clustered.
Further, we find that
half of our YSO  population is located in regions denser than
1000~YSOs/deg$^2$ ($\sim$10$^5$~YSOs/pc$^2$) and cover a relatively small fraction of the sky ($\sim$0.5
deg$^2$ compared to the $\sim$9 deg$^2$ observed).

\begin{figure}\centering 
\plotone{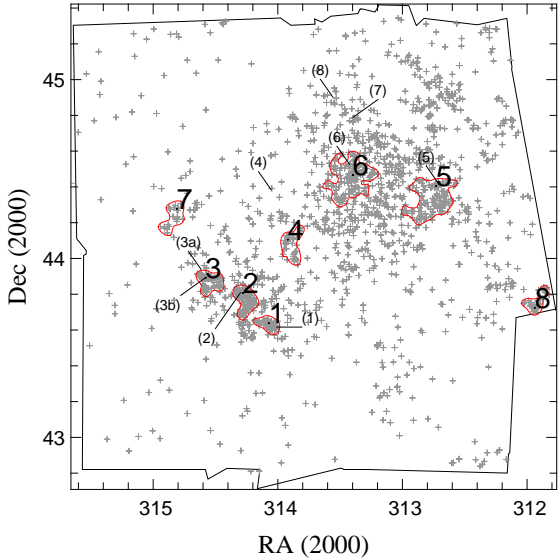}
\caption{Location of clusters shown on the distribution of
  the $\sim$1600 YSO candidates (plus symbols). Solid red lines are contours of clusters we adopted
  (see text for more details). Our eight clusters are labelled with large, bold numbers and 
  no brackets.
  The location of clusters found
  by \cite{Cambresy-2002} are also indicated by smaller bracketed
  labels.} 
  \label{fig:clusters}
\end{figure}

From the distribution of YSOs,  we distinguished by eye 8 main
clusters (for a discussion of more quantitative means to identify
clusters of YSOs from similar catalogs, see for example Jorgensen et al. 2008). 
We determined the center of each cluster as the local peak of the YSO density. 
We arbitrarily define the boundaries of these clusters as the
YSO density contour at a level of 1/4 of the maximum density peak. 
The location, the size, the
number of stars and the mean $A_v$ for each 
cluster is provided in Table
\ref{tab:clusters}. Five of these clusters have been previously identified  by
\cite{Cambresy-2002}; they are the most prominent clusters located
inside  the Gulf of Mexico and in the central part of the NANeb. 
Figure~\ref{fig:clusters} shows the location of both our clusters
and clusters defined by Cambr\'esy et al. One can see in this figure that 
cluster \#4 of \cite{Cambresy-2002} does not exist in our YSO candidate
distribution. Clusters \#7 and \#8 of Cambresy appear in our YSO
distribution, but are similar to many other small clusters in the NANeb; 
they are not as densely populated as the ones we have defined. 
Because the \cite{Cambresy-2002} technique is sensitive to both
WTTs and CTTs, whereas ours essentially excludes WTTs, it is reasonable
that our two methods will yield somewhat different results.  It is
likely that the clusters found by \cite{Cambresy-2002} but not by
us may be the evolutionarily older clusters in the region.

\begin{deluxetable}{ccccccc}
\tablecaption{NANeb clusters \label{tab:clusters}}
\tablehead{id&  RA         & Dec        & Area   & YSOs  & $<A_v>$ & idc     \\
  & J(2000)     & J(2000)    & deg$^2$& Count & mag}
\startdata
1 &20 56 17.4 & +43 38 18.8 & 0.015 &    48 &    18.12  & 1  \\  
2 &20 57 07.1 & +43 48 21.8 & 0.025 &   120 &    16.47  & 2  \\  
3 &20 58 19.0 & +43 53 32.5 & 0.023 &    74 &     4.93  & 3b \\  
4 &20 55 40.5 & +44 06 20.0 & 0.025 &    39 &     5.68  & \dots   \\  
5 &20 50 54.9 & +44 24 18.1 & 0.067 &   121 &     5.54  & 5  \\  
6 &20 53 35.2 & +44 27 57.4 & 0.075 &   130 &     6.27  & 6  \\  
7 &20 59 14.2 & +44 16 41.3 & 0.022 &    22 &     1.48  & \dots \\  
8 &20 47 45.1 & +43 43 29.4 & 0.017 &    28 &     \dots & \dots
\enddata
\tablenotetext{}{The id, center coordinate and area of identified clusters are
 indicated on the first 4 columns, followed by the number of YSO members 
 and the average reddening from the extinction map of
 \cite{Cambresy-2002}. 
 The last column is the cluster label from \cite{Cambresy-2002}, if appropriate. }
\end{deluxetable}

\begin{figure}
\centering
\epsscale{0.8}
\plotone{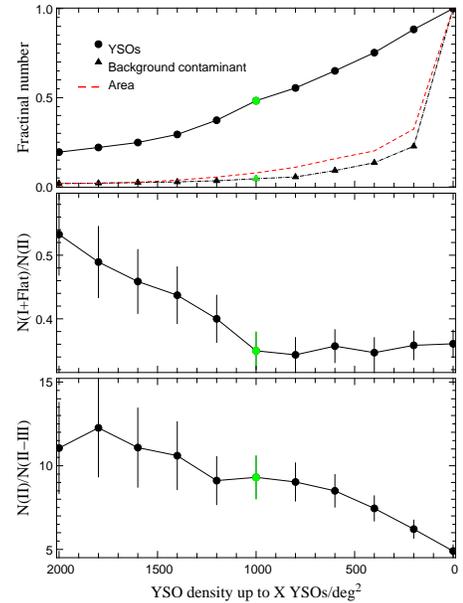}
\caption{Ratios of object classes as a function of YSO density.
The $x$-axis corresponds to the YSO density contours plotted in Figure
\ref{fig:distrib} from hight density to low density (corresponding to small and large 
surface surrounded by each contour on Figure \ref{fig:distrib}). 
Inside each contour, we calculated several fractional  
numbers:  TOP: the fraction of YSO candidates (filled dot symbols linked by solid line) 
and  background contaminants (filled triangle liked by dot-dashed line); 
the red dashed line shows the surface area surrounded
by each contour as a fraction of the total covered surface ($\sim$9 deg$^2$)
MIDDLE: same, but for the fraction of (Class~I+Flat) / Class~II.
BOTTOM: same, but for the fraction of Class~II / Class~\III. 
Note that for the last value, $X$=0 corresponds physically to the boundary of our IRAC coverage.
}
\label{fig:evolR}
\epsscale{1}
\end{figure}

In the bottom panel of Figure \ref{fig:distrib}, we plot the
location of YSOs with different symbols corresponding to their class. To
investigate in more detail a possible spatial segregation between
Class~I(+flat), II, and \III, we used the YSO density contours to define
regions within which   
we have computed the ratio of  N(Class~I + Flat) / N(Class~II) and
N(Class~II) / N(Class~\III). 
The behavior of these ratios as a function of the density contour limit 
appears in Figure \ref{fig:evolR}. This Figure shows that in dense
regions (i.e., more clustered regions), the proportion of Class\,II
objects
compared to the (presumably) younger Class~I's is statistically lower than in the entire
observed region by a factor of $\sim$1.3. 
This is consistent with expectations - either due to high velocity
stars moving away from their birth sites, or as a result of the
overall expansion of young clusters as they age due to,
for example, removal of their remaining molecular cloud material
by stellar winds.     
Moreover, the ratio N(Class~II)/N(Class~\III) decreases from denser to
less dense regions, as expected for the same reasons, by a factor of
$\sim$2.

Figure \ref{fig:evolR} also shows the
fractional number of background  
contaminants as compared to the fractional surface area inside
each of the YSO density contours.  
The proportional number of background contaminants follows very
closely the proportional surface area. This suggests that sources
flagged as contaminants are really background objects and not YSOs;
otherwise
one would expect an excess of sources flagged as background inside
dense regions. 

\begin{figure}\gpos
\plotone{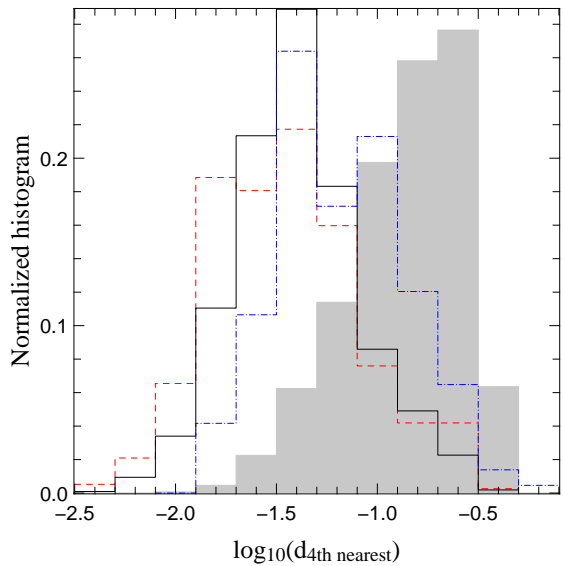}
\caption {Histogram of the log of  the distance of
the 4th nearest YSO (in degrees) for each of our YSO candidates. The red dashed
line is for Class~I+Flat spectrum sources, the solid black line for Class~II stars
and the blue dash-dot histogram for Class~\III. The gray histogram
represents the remaining population of non-cluster members. 
The Class~I population, usually
assumed to be younger, is significantly more clustered than
Class~\III; this is consistent with expectations that, as stars
dynamically evolve, they move away from their birthplace.} 
\label{fig:dist_tennearest}
\end{figure}

Another way to characterize the degree of clustering, used by
many authors, is the distribution of nearest neighbors.  
For each of our YSO candidates, we calculated the distance to
the 4th nearest YSO candidates; Figure~\ref{fig:dist_tennearest}
shows the histogram of this  distance, displayed for
Class~I+Flat, Class~II and Class~\III. We find a statistically
significant difference between the 3 histograms: using a
Kolmogorov-Smirnov 
test, the probability that each of these distributions is
drawn from the same parent distribution as either of the other two
distributions is less than 0.02\%.  
This is consistent with what 
we found above: the Class~I population, usually
assumed to be younger, is significantly more clustered than
Class~\III. 

\section{HH Objects}
\label{sec:hhobj}

\cite{Ogura-2002} identified some HH objects using H$\alpha$ over a
small region of the complex near bright-rimmed cloud (BRC) 31.
\cite{Bally-2003} studied the entire Pelican Nebula using
H$\alpha$, [\ion{N}{2}], and [\ion{S}{2}], finding several new HH
objects.  

The 3.5 to 8.5\,$\mu$m wavelength range covered by the IRAC four channels is 
quite suitable to study deeply embedded young stellar outflows, since this range
includes some of the brightest collisionally excited pure rotational H$_2$ 
emission lines \citep{Wright-1996,Noriega-Crespo-2004a,Noriega-Crespo-2004b}  as well 
as other H$_2$ vibrational lines \citep{Smith-2005}. At the longer 
wavelengths, the MIPS channels sample a rich combination of atomic/ionic
and molecular lines that include some of the best atomic species to study
outflows, e.g. [Fe~II] 25.98\,$\mu$m, [O~I] 63.18\,$\mu$m and [C~II] 157.74\,$\mu$m
(see e.g. \citealt{Molinari-2000,Morris-2004}) that fall within the 24, 
70 and 160\,$\mu$m bandpasses, respectively, plus of course continuum
emission  from cold dust. We discuss below  our observations of one HH
object discovered previously, HH~555.

HH~555 itself is not an embedded flow (like e.g. HH~211) since it is 
clearly detected at optical wavelengths (H$\alpha$ and [S~II] 6717/31\,\AA)
and its spectrum is consistent with that of a shock excited gas moving 
supersonically \citep{Bally-2003}. HH~555 belongs to that special class
of irradiated jets, like HH~399 in the Trifid nebula
\citep{Yusef-Zadeh-2005}  that are found in active star forming
regions and are surrounded by a bath of UV ionizing photons from
recently formed massive stars.

\begin{figure*}
\plotone{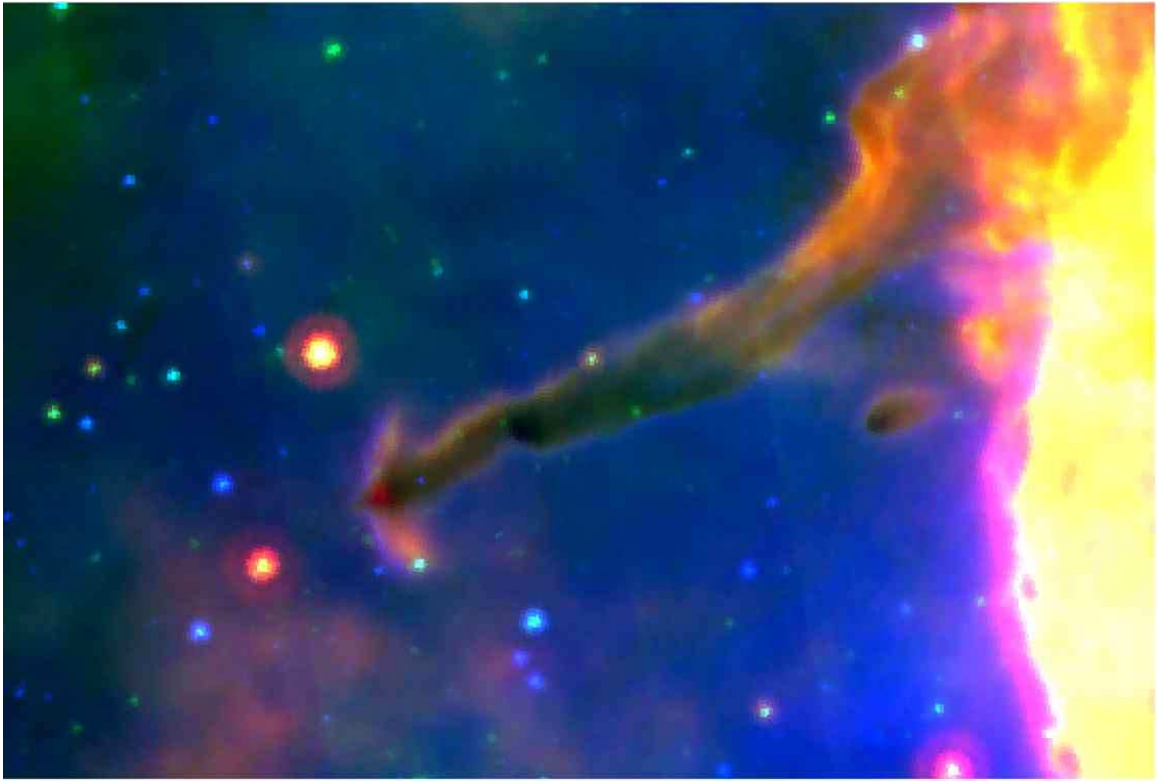}
\caption{Three color image of HH~555 arising from the tip of a dense pillar using 
H$\alpha$ (blue), IRAC 8\,$\mu$m (green) and MIPS 24\,$\mu$m (red). 
The FOV is $\sim 6\arcmin \times 4\arcmin$, and  North is up and East is 
left. The mid/far infrared bipolar jet ($\sim$ 1\arcmin~in length)
is centered on a 24\,$\mu$m~ embedded source. The stratification in ionization
along the jet from East-to-West (the gradient in color) can be explained 
by the shield created during the two (stellar and photoevaporation) wind 
interaction \citep{Kajdic-2007}.}
\label{fig:ha+8+24}
\end{figure*}

The case for HH~555 is particularly interesting because the jet and counter-jet
are bent westward as soon as they arise from their embedded source  at
the dense tip of the pillar, indicating the presence of stellar ``sidewind'' 
\citep{Masciadri-2001}. The pillar itself is being photoevaporated 
leading to the formation of a two-shock structure (see \citealt{Kajdic-2007} 
for details). 
The length of time that the jet will maintain its bipolar morphology
is expected to depend on the strength of the impinging ionizing flux.
The two-shock structure developed by the interacting winds actually 
creates a shield that protects both the jet and the pillar, and therefore, 
one would expect lower ionization on the West side of the outflow. 
Indeed, the Spitzer images combined with the ground based H$\alpha$
emission (Figure \ref{fig:ha+8+24})
confirms this stratification, if one assumes that the 8\,$\mu$m and 
24\,$\mu$m~emission arises mostly from H$_2$ and [S~II], as is the case in 
other jets. This also explains why  the 24\,$\mu$m source lies in the middle
of the 8\,$\mu$m~and 24\,$\mu$m~jets, but is slightly offset with respect 
to H$\alpha$ (see e.g. \citealt{Kajdic-2007}, Fig.~3).

Our 70\,$\mu$m images do not show a clear signature of the jet, and this
could be due to a combination of things: lack of sensitivity, artifacts along
the scanning direction and physics ([O~I] 63.18\,$\mu$m is not present, neither
is cold dust). Careful analysis of the data do show the presence of the 
source at 70\,$\mu$m~at the tip of the pillar. The pillar itself, although 
faint, is distinguished and its emission is likely to be due to [O~I] 
photodissociated emission. We measured the flux at 24 and 70\,$\mu$m
using a small aperture of 7\arcsec\ and 16\arcsec\ respectively, centered at
20h51m19.52s +44d25m38.3s. We found 10$\pm$3~mJy at 24\,$\mu$m and
830$\pm$220~mJy at 70\,$\mu$m.

\section{Conclusion}

We have combined mid-IR photometry from Spitzer/IRAC with $JHK_s$ data
from 2MASS to provide a sensitive photometric census of objects
towards the North America and Pelican Nebulae.
We have used IRAC color and magnitude diagnostics to identify more
than 1600 sources with infrared excess characteristic of young
stars surrounded by a disk.

Our YSO candidates are located primarily on the central dark cloud
(Lynds 935). We identified 8 main clusters,  which together contain
about a third of all the YSOs in the region; 5 of these clusters
were previously recognized by \cite{Cambresy-2002}. The sources we
identify as background galaxies or AGN are relatively uniformly
spread across the observed region.  These two properties suggest
a low level of contamination in our YSO sample.

We have assigned our YSOs classifications in the Class\,I/Flat/II/III system
according to
their infrared SED slope. The proportion of Class~I+Flat compared to
Class~II is variable within the NANeb, with the dark clouds called
``Gulf of Mexico'' having a remarkably higher proportion of Class~I's,
suggesting these clusters have
a younger age than the rest of NANeb. The ``Gulf of Mexico'' region
will be discussed in more detail in R09 where the MIPS data for
NANeb will be presented.

We compared optical photometry of a small sample of the YSO candidates to
evolutionary models in order to provide an age estimate for the
NANeb. We found that most of these YSOs appear younger than 10\,Myr according
to the isochrones of \cite{Siess-2000}, with the median
age being 3\,Myr. However, the most embedded and probably youngest
YSOs were not included in this sample because our optical photometry is
less sensitive than our IRAC observations  and because the youngest
stars are very red. However, despite all the caveats regarding this age
estimation, it is the most robust published so far.

Finally, thanks to our infrared images from 3.6 to 70\,$\mu$m, we have
provided new
clues on the nature of HH~555, a peculiar HH object reported in
\cite{Bally-2003}. We detected a source at 70\,$\mu$m at the
base of the bi-polar jets.  Our data add support to the hypothesis
that the bent shape of HH~555 is a result of having the jet propagate
in an environment with a stellar ``sidewind''.

\acknowledgements 
This work is based in part on observations made with the Spitzer Space 
Telescope, which is operated by the Jet Propulsion Laboratory,
California Institute of Technology under a contract with NASA. Support 
for this work was provided by NASA through an award issued by JPL/Caltech.
This research has made use of NASA's Astrophysics Data System (ADS)
Abstract Service, and of the SIMBAD database, operated at CDS,
Strasbourg, France.  This research has made use of data products from
the Two Micron All-Sky Survey (2MASS), which is a joint project of the
University of Massachusetts and the Infrared Processing and Analysis
Center, funded by the National Aeronautics and Space Administration
and the National Science Foundation.  These data were served by the
NASA/IPAC Infrared Science Archive, which is operated by the Jet
Propulsion Laboratory, California Institute of Technology, under
contract with the National Aeronautics and Space Administration.  This
research has made use of the Digitized Sky Surveys, which were
produced at the Space Telescope Science Institute under U.S.
Government grant NAG W-2166. The images of these surveys are based on
photographic data obtained using the Oschin Schmidt Telescope on
Palomar Mountain and the UK Schmidt Telescope. The plates were
processed into the present compressed digital form with the permission
of these institutions.

The research described in this paper was partially carried out at the
Jet Propulsion Laboratory, California Institute of Technology, under
contract with the National Aeronautics and Space Administration.


\begin{thebibliography}{36}
\expandafter\ifx\csname natexlab\endcsname\relax\def\natexlab#1{#1}\fi

\bibitem[{{Andr{\'e}}(1994)}]{Andre-1994}
{Andr{\'e}}, P. 1994, in The Cold Universe, ed. T.~{Montmerle}, C.~J. {Lada},
  I.~F. {Mirabel}, \& J.~{Tran Thanh van}, 179

\bibitem[{{Armond} {et~al.}(2003){Armond}, {Reipurth}, \& {Vaz}}]{armond-2003}
{Armond}, A.~C., {Reipurth}, B., \& {Vaz}, L.~P. 2003, Bulletin of the
  Astronomical Society of Brazil, 23, 92

\bibitem[{{Bally} \& {Reipurth}(2003)}]{Bally-2003}
{Bally}, J. \& {Reipurth}, B. 2003, \aj, 126, 893

\bibitem[{{Bally} \& {Scoville}(1980)}]{Bally-1980}
{Bally}, J. \& {Scoville}, N.~Z. 1980, \apj, 239, 121

\bibitem[{{Baraffe} {et~al.}(1998){Baraffe}, {Chabrier}, {Allard}, \&
  {Hauschildt}}]{Baraffe-1998}
{Baraffe}, I., {Chabrier}, G., {Allard}, F., \& {Hauschildt}, P.~H. 1998, \aap,
  337, 403

\bibitem[{{Cambr{\'e}sy} {et~al.}(2002){Cambr{\'e}sy}, {Beichman}, {Jarrett},
  \& {Cutri}}]{Cambresy-2002}
{Cambr{\'e}sy}, L., {Beichman}, C.~A., {Jarrett}, T.~H., \& {Cutri}, R.~M.
  2002, \aj, 123, 2559

\bibitem[{{Comer{\'o}n} \& {Pasquali}(2005)}]{Comeron-2005b}
{Comer{\'o}n}, F. \& {Pasquali}, A. 2005, \aap, 430, 541

\bibitem[{{Fazio} {et~al.}(2004){Fazio}, {Hora}, {Allen}, {Ashby}, {Barmby},
  {Deutsch}, {Huang}, {Kleiner}, {Marengo}, {Megeath}, {Melnick}, {Pahre},
  {Patten}, {Polizotti}, {Smith}, {Taylor}, {Wang}, {Willner}, {Hoffmann},
  {Pipher}, {Forrest}, {McMurty}, {McCreight}, {McKelvey}, {McMurray}, {Koch},
  {Moseley}, {Arendt}, {Mentzell}, {Marx}, {Losch}, {Mayman}, {Eichhorn},
  {Krebs}, {Jhabvala}, {Gezari}, {Fixsen}, {Flores}, {Shakoorzadeh}, {Jungo},
  {Hakun}, {Workman}, {Karpati}, {Kichak}, {Whitley}, {Mann}, {Tollestrup},
  {Eisenhardt}, {Stern}, {Gorjian}, {Bhattacharya}, {Carey}, {Nelson},
  {Glaccum}, {Lacy}, {Lowrance}, {Laine}, {Reach}, {Stauffer}, {Surace},
  {Wilson}, {Wright}, {Hoffman}, {Domingo}, \& {Cohen}}]{Fazio-2004}
{Fazio}, G.~G., {Hora}, J.~L., {Allen}, L.~E., {et~al.} 2004, \apjs, 154, 10

\bibitem[{{Flaherty} {et~al.}(2007){Flaherty}, {Pipher}, {Megeath}, {Winston},
  {Gutermuth}, {Muzerolle}, {Allen}, \& {Fazio}}]{Flaherty-2007}
{Flaherty}, K.~M., {Pipher}, J.~L., {Megeath}, S.~T., {et~al.} 2007, \apj, 663,
  1069

\bibitem[{{Gutermuth} {et~al.}(2008){Gutermuth}, {Myers}, {Megeath}, {Allen},
  {Pipher}, {Muzerolle}, {Porras}, {Winston}, \& {Fazio}}]{Gutermuth-2008a}
{Gutermuth}, R.~A., {Myers}, P.~C., {Megeath}, S.~T., {et~al.} 2008, \apj, 674,
  336

\bibitem[{{Harvey} {et~al.}(2007{\natexlab{a}}){Harvey}, {Mer{\'{\i}}n},
  {Huard}, {Rebull}, {Chapman}, {Evans}, \& {Myers}}]{Harvey-2007b}
{Harvey}, P., {Mer{\'{\i}}n}, B., {Huard}, T.~L., {et~al.} 2007{\natexlab{a}},
  \apj, 663, 1149

\bibitem[{{Harvey} {et~al.}(2006){Harvey}, {Chapman}, {Lai}, {Evans}, {Allen},
  {J{\o}rgensen}, {Mundy}, {Huard}, {Porras}, {Cieza}, {Myers}, {Mer{\'{\i}}n},
  {van Dishoeck}, {Young}, {Spiesman}, {Blake}, {Koerner}, {Padgett},
  {Sargent}, \& {Stapelfeldt}}]{Harvey-2006}
{Harvey}, P.~M., {Chapman}, N., {Lai}, S.-P., {et~al.} 2006, \apj, 644, 307

\bibitem[{{Harvey} {et~al.}(2007{\natexlab{b}}){Harvey}, {Rebull}, {Brooke},
  {Spiesman}, {Chapman}, {Huard}, {Evans}, {Cieza}, {Lai}, {Allen}, {Mundy},
  {Padgett}, {Sargent}, {Stapelfeldt}, {Myers}, {van Dishoeck}, {Blake}, \&
  {Koerner}}]{Harvey-2007a}
{Harvey}, P.~M., {Rebull}, L.~M., {Brooke}, T., {et~al.} 2007{\natexlab{b}},
  \apj, 663, 1139

\bibitem[{{Herbig}(1958)}]{herbig-1958}
{Herbig}, G.~H. 1958, \apj, 128, 259

\bibitem[{{Jeffries} {et~al.}(2007){Jeffries}, {Oliveira}, {Naylor}, {Mayne},
  \& {Littlefair}}]{Jeffries-2007}
{Jeffries}, R.~D., {Oliveira}, J.~M., {Naylor}, T., {Mayne}, N.~J., \&
  {Littlefair}, S.~P. 2007, \mnras, 376, 580

\bibitem[{{Kajdi{\v c}} \& {Raga}(2007)}]{Kajdic-2007}
{Kajdi{\v c}}, P. \& {Raga}, A.~C. 2007, \apj, 670, 1173

\bibitem[{{Lada}(1987)}]{Lada-1987}
{Lada}, C.~J. 1987, in IAU Symp. 115: Star Forming Regions, ed. M.~{Peimbert}
  \& J.~{Jugaku}, 1--17

\bibitem[{{Laugalys} \& {Strai{\v z}ys}(2002)}]{laugalys-2002}
{Laugalys}, V. \& {Strai{\v z}ys}, V. 2002, Baltic Astronomy, 11, 205

\bibitem[{{Luhman} {et~al.}(2003){Luhman}, {Stauffer}, {Muench}, {Rieke},
  {Lada}, {Bouvier}, \& {Lada}}]{Luhman-2003b}
{Luhman}, K.~L., {Stauffer}, J.~R., {Muench}, A.~A., {et~al.} 2003, \apj, 593,
  1093

\bibitem[{{Makovoz} \& {Marleau}(2005)}]{Makovoz-2005}
{Makovoz}, D. \& {Marleau}, F.~R. 2005, \pasp, 117, 1113

\bibitem[{{Masciadri} \& {Raga}(2001)}]{Masciadri-2001}
{Masciadri}, E. \& {Raga}, A.~C. 2001, \aj, 121, 408

\bibitem[{{Molinari} {et~al.}(2000){Molinari}, {Noriega-Crespo}, {Ceccarelli},
  {Nisini}, {Giannini}, {Lorenzetti}, {Caux}, {Liseau}, {Saraceno}, \&
  {White}}]{Molinari-2000}
{Molinari}, S., {Noriega-Crespo}, A., {Ceccarelli}, C., {et~al.} 2000, \apj,
  538, 698

\bibitem[{{Morris} {et~al.}(2004){Morris}, {Noriega-Crespo}, {Marleau},
  {Teplitz}, {Uchida}, \& {Armus}}]{Morris-2004}
{Morris}, P.~W., {Noriega-Crespo}, A., {Marleau}, F.~R., {et~al.} 2004, \apjs,
  154, 339

\bibitem[{{Muench} {et~al.}(2007){Muench}, {Lada}, {Luhman}, {Muzerolle}, \&
  {Young}}]{Muench-2007}
{Muench}, A.~A., {Lada}, C.~J., {Luhman}, K.~L., {Muzerolle}, J., \& {Young},
  E. 2007, \aj, 134, 411

\bibitem[{{Noriega-Crespo} {et~al.}(2004{\natexlab{a}}){Noriega-Crespo},
  {Moro-Martin}, {Carey}, {Morris}, {Padgett}, {Latter}, \&
  {Muzerolle}}]{Noriega-Crespo-2004a}
{Noriega-Crespo}, A., {Moro-Martin}, A., {Carey}, S., {et~al.}
  2004{\natexlab{a}}, \apjs, 154, 402

\bibitem[{{Noriega-Crespo} {et~al.}(2004{\natexlab{b}}){Noriega-Crespo},
  {Morris}, {Marleau}, {Carey}, {Boogert}, {van Dishoeck}, {Evans}, {Keene},
  {Muzerolle}, {Stapelfeldt}, {Pontoppidan}, {Lowrance}, {Allen}, \&
  {Bourke}}]{Noriega-Crespo-2004b}
{Noriega-Crespo}, A., {Morris}, P., {Marleau}, F.~R., {et~al.}
  2004{\natexlab{b}}, \apjs, 154, 352

\bibitem[{{Ogura} {et~al.}(2002){Ogura}, {Sugitani}, \& {Pickles}}]{Ogura-2002}
{Ogura}, K., {Sugitani}, K., \& {Pickles}, A. 2002, \aj, 123, 2597

\bibitem[{{Rebull} {et~al.}(2002){Rebull}, {Makidon}, {Strom}, {Hillenbrand},
  {Birmingham}, {Patten}, {Jones}, {Yagi}, \& {Adams}}]{Rebull-2002}
{Rebull}, L.~M., {Makidon}, R.~B., {Strom}, S.~E., {et~al.} 2002, \aj, 123,
  1528

\bibitem[{{Siess} {et~al.}(2000){Siess}, {Dufour}, \& {Forestini}}]{Siess-2000}
{Siess}, L., {Dufour}, E., \& {Forestini}, M. 2000, \aap, 358, 593

\bibitem[{{Silverman}(1986)}]{Silverman-1986}
{Silverman}, B.~W. 1986, {Density estimation for statistics and data analysis}
  (Monographs on Statistics and Applied Probability, London: Chapman and Hall,
  1986)

\bibitem[{{Skrutskie} {et~al.}(2006){Skrutskie}, {Cutri}, {Stiening},
  {Weinberg}, {Schneider}, {Carpenter}, {Beichman}, {Capps}, {Chester},
  {Elias}, {Huchra}, {Liebert}, {Lonsdale}, {Monet}, {Price}, {Seitzer},
  {Jarrett}, {Kirkpatrick}, {Gizis}, {Howard}, {Evans}, {Fowler}, {Fullmer},
  {Hurt}, {Light}, {Kopan}, {Marsh}, {McCallon}, {Tam}, {Van Dyk}, \&
  {Wheelock}}]{Skrutskie-2006}
{Skrutskie}, M.~F., {Cutri}, R.~M., {Stiening}, R., {et~al.} 2006, \aj, 131,
  1163

\bibitem[{{Smith} \& {Rosen}(2005)}]{Smith-2005}
{Smith}, M.~D. \& {Rosen}, A. 2005, \mnras, 357, 1370

\bibitem[{{Stauffer}(1996)}]{Stauffer-1996}
{Stauffer}, J.~R. 1996, in Astronomical Society of the Pacific Conference
  Series, Vol. 109, Cool Stars, Stellar Systems, and the Sun, ed.
  R.~{Pallavicini} \& A.~K. {Dupree}, 305

\bibitem[{{Werner} {et~al.}(2004){Werner}, {Roellig}, {Low}, {Rieke}, {Rieke},
  {Hoffmann}, {Young}, {Houck}, {Brandl}, {Fazio}, {Hora}, {Gehrz}, {Helou},
  {Soifer}, {Stauffer}, {Keene}, {Eisenhardt}, {Gallagher}, {Gautier}, {Irace},
  {Lawrence}, {Simmons}, {Van Cleve}, {Jura}, {Wright}, \&
  {Cruikshank}}]{Werner-2004}
{Werner}, M.~W., {Roellig}, T.~L., {Low}, F.~J., {et~al.} 2004, \apjs, 154, 1

\bibitem[{{Wright} {et~al.}(1996){Wright}, {Drapatz}, {Timmermann}, {van der
  Werf}, {Katterloher}, \& {de Graauw}}]{Wright-1996}
{Wright}, C.~M., {Drapatz}, S., {Timmermann}, R., {et~al.} 1996, \aap, 315,
  L301

\bibitem[{{Yusef-Zadeh} {et~al.}(2005){Yusef-Zadeh}, {Biretta}, \&
  {Wardle}}]{Yusef-Zadeh-2005}
{Yusef-Zadeh}, F., {Biretta}, J., \& {Wardle}, M. 2005, \apj, 624, 246

\end{thebibliography}

\clearpage
\begin{landscape}
\begin{deluxetable*}{cccccccccccccccccccc}
\tabletypesize{\scriptsize}
\tablecolumns{20}
\tablecaption{Table of YSO candidates\tablenotemark{a}
\label{tab:ysocand}}
\tablehead{\colhead{IAU Designation} & \colhead{RA (2000)} &
  \colhead{   Dec (2000)} & \colhead{        2MASS id} & \colhead{  $B$} & \colhead{   $V$} & \colhead{   $I$} & \colhead{   $J$} & \colhead{   $H$} & \colhead{  $K_s$} & \colhead{ $[3.6]$} & \colhead{ $[4.5]$} & \colhead{ $[5.8]$} & \colhead{ $[8.0]$} & \colhead{  $[24]$}
}
\startdata
                      SST205659.3+434753.0 &            20 56 59.326  &            43 47 52.993  &            20565934+4347530 &            20.195 &            18.747 &            15.963 &            14.286 &            13.485 &            13.174 &            12.609 &            12.287 &            11.892 &            11.348 &            \dots\\
                      SST205646.8+434609.0 &            20 56 46.765  &            43 46 08.980  &            20564675+4346089 &            19.981 &            18.023 &            15.385 &            13.226 &            12.180 &            11.783 &            11.653 &            11.485 &            11.244 &            10.392 &             5.060\\
\enddata
\tablenotetext{a}{Table will be presented in its entirety in the
electronic version.  Headings are shown here for guidance regarding
its content.}
\end{deluxetable*}
 \clearpage
\end{landscape}

\end{document}